%% file: ms_mh.tex
\documentclass[useAMS,usenatbib]{mn2e}

\usepackage{times}  
\usepackage{listings}
\usepackage{graphics}
\usepackage{subfigure}
\usepackage{graphicx}
\usepackage{graphicx,xspace}
\usepackage{amssymb}
\usepackage{amsmath}
\usepackage{hyperref}
\usepackage{aas_macros}
\usepackage{lscape}
\usepackage{rotating}
\usepackage{placeins}
\usepackage{natbib}
\usepackage{color}

\newcommand{\galform}{{\sc{galform}}\xspace}

\title[The evolution of the SHM relationship]
{The evolution of the stellar mass versus halo mass relationship}
\author[Peter D. Mitchell]{
\parbox[t]{\textwidth}{
Peter D. Mitchell\thanks{\rm E-mail: peter.mitchell@durham.ac.uk }, 
Cedric G. Lacey,
Carlton M. Baugh,
Shaun Cole
}
\vspace*{6pt} \\
Institute for Computational Cosmology, Department of Physics,
University of Durham, South Road, Durham, DH1 3LE, UK.
\vspace*{-0.5cm}}
       
\begin{document}
\date{\today}
\pagerange{\pageref{firstpage}--\pageref{lastpage}} \pubyear{2013}
\maketitle
\label{firstpage}

\begin{abstract}
We present an analysis of the predictions made by the \galform semi-analytic galaxy formation model for the evolution of the relationship between stellar mass and halo mass.
We show that for the standard implementations of supernova feedback and gas reincorporation used in semi-analytic models, this relationship is predicted to evolve weakly over the redshift range $0<z<4$.
Modest evolution in the median stellar mass versus halo mass (SHM) relationship implicitly requires that, at fixed halo mass, the efficiency of stellar mass assembly must be almost constant with cosmic time.
We show that in our model, this behaviour can be understood in simple terms as a result of a constant efficiency of gas reincorporation, and an efficiency of SNe feedback that is, on average, constant at fixed halo mass.
We present a simple explanation of how feedback from active galactic nuclei (AGN) acts in our model to introduce a break in the SHM relation whose location is predicted to evolve only modestly.
Finally, we show that if modifications are introduced into the model such that, for example, the gas reincorporation efficiency is no longer constant, the median SHM relation is predicted to evolve significantly over $0<z<4$.
Specifically, we consider modifications that allow the model to better reproduce either the evolution of the stellar mass function or the evolution of average star formation rates inferred from observations.
\end{abstract}

\begin{keywords}
galaxies: formation -- galaxies: evolution -- galaxies: haloes -- galaxies: stellar content
\end{keywords}

\section{Introduction}
\label{Introduction_HM}

Over the last decade, interest has grown in using statistical inference to construct empirical
models that describe how galaxies are distributed within dark matter haloes  
\cite[e.g.][]{Peacock00,Scoccimarro01,Berlind02,Vale04,Wang13,Lu14b}.
Observational constraints for these models typically include a selection of measurements of the 
abundances, clustering and lensing of galaxies, which are then combined with theoretical predictions 
for the abundance and clustering of dark matter haloes. Earlier work in this area typically used 
galaxy abundances and/or clustering as a function of luminosity to constrain model parameters 
\cite[e.g.][]{Berlind02, Yang03, Conroy06}. As multi-wavelength galaxy surveys have become 
available, it has become commonplace to replace galaxy luminosity with stellar mass (which can be 
estimated from broad-band photometry) as the dependent variable in this type of analysis 
\cite[e.g.][]{Mandelbaum06, Behroozi10, Guo10, Moster10}. It has also become possible to place constraints
on the relationship between galaxies and haloes for redshifts up to $z=1$ and beyond
\cite[e.g.][]{Wake11, Behroozi13, Moster13, Shankar14, Verlander14, Durkalec15, McCracken15}.

\begin{figure*}
\begin{center}
\includegraphics[width=40pc]{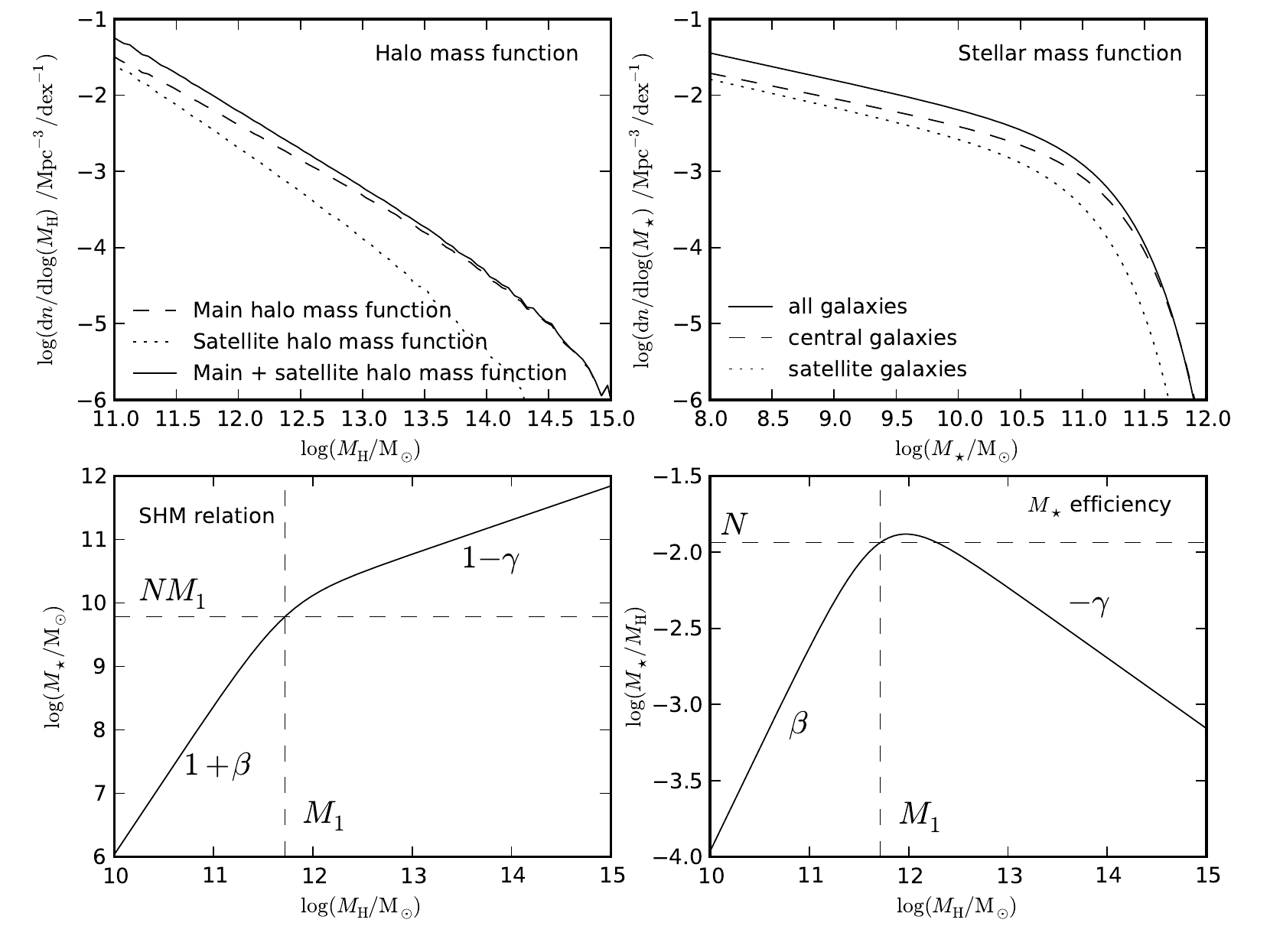}
\caption{
Schematic (based on the reference model detailed in Section~\ref{GALFORM_Section_HM}) to demonstrate the relationship at $z=0$ between the halo mass function, the stellar mass function and the median SHM relation.
The meanings of the parameters from Eqn.~\ref{shm_eqn} are also illustrated. 
For example, in the lower right panel, the dashed vertical line shows the SHM break mass, $M_{\mathrm{1}}$, and the dashed horizontal line shows the SHM normalisation, $N$.
$\beta$ and $-\gamma$ are the power law slopes shown below and above the SHM break respectively.
{\it Upper left:} Main halo mass function (dashed), satellite halo mass function (dotted) and combined main plus satellite halo mass function (solid). 
The satellite halo mass plotted is the mass of the host subhalo at infall.
{\it Upper right:} Stellar mass function of all galaxies (solid), central galaxies (dashed) and satellite galaxies (dotted).
{\it Lower left:} Stellar mass versus halo mass (SHM) relation.
{\it Lower right:} Median stellar mass assembly efficiency, $M_{\star}/M_{\mathrm{H}}$, plotted as a function of halo mass.}
\label{cartoon}
\end{center}
\end{figure*}

A strong consensus that has emerged from studies of this type is that the dependence of median 
galaxy stellar mass, $M_\star$, on halo mass, $M_{\mathrm{H}}$, (hereafter referred to as the SHM relation) 
can be simply described by two power laws that connect at a stellar mass that corresponds roughly 
to the knee of the stellar mass function \cite[e.g.][]{Moster10, Yang12}. While more complex 
parametrisations have been advocated \cite[e.g.][]{Behroozi10, Behroozi13}, the basic picture is 
that there are two regimes (the two power laws) that describe how the relative 
efficiency\footnote{We will refer to $M_\star / M_{\mathrm{H}}$ as an ``efficiency'', even though
more correctly this is given by $M_\star / (f_{\mathrm{B}} M_{\mathrm{H}})$, where $f_{\mathrm{B}}$
is the universal baryon fraction.} of stellar mass 
assembly\footnote{We use the convention that stellar mass assembly refers to both star formation within
a galaxy and to stellar mass brought in by galaxy mergers.}, $M_\star / M_{\mathrm{H}}$, drops away either side of a peak 
value at the halo mass where the two power laws meet.
An illustration of the relationship between the 
halo mass function, the stellar mass function and the median SHM relation
is shown in Fig.~\ref{cartoon}.

Arguably, a weaker level of consensus has been achieved regarding the amount of evolution in the median SHM relation that is implied
by observational data. For example, \cite{Behroozi13} report that the SHM relation is marginally consistent
with no evolution over the range $0 < z< 6$, although their analysis prefers a solution where the SHM break halo mass
evolves, peaking at $z=2$. Their results also show little evidence for a significant
variation in the peak stellar mass assembly efficiency\footnote{By this we mean the maximum value of $M_\star/M_{\mathrm{H}}$.} for $z<4$. The analysis of \cite{Moster13} instead finds significant evidence for monotonic evolution in all 
SHM relation parameters (including the SHM break mass) over $0<z<4$, and that the peak stellar mass assembly efficiency
also evolves significantly over this redshift range. Over a more limited redshift range ($0.2<z<1$), \cite{Leauthaud12} report
that the SHM break mass increases but that the peak stellar mass assembly efficiency remains constant, consistent
with \cite{Behroozi13}. In contrast to these three studies, \cite{Hudson15} (for $0.2<z<0.8$) and 
\cite{McCracken15} (for $0.5<z<2$) report that the SHM break mass is constant over their respective redshift
intervals. \cite{Hudson15} also find that the peak stellar mass assembly efficiency does evolve significantly
over $0.2<z<0.8$, in agreement with \cite{Moster13}.

Disagreements between different studies are not surprising for a number of reasons. One possible source of
error can be attributed to uncertain stellar mass estimates which can strongly affect 
the inferred stellar mass function, particularly for massive galaxies \cite[e.g.][]{Marchesini09,Behroozi10,Mitchell13}. 
Fairly strong priors on the distribution of errors on stellar mass estimates have to be adopted 
when constraining SHM parameters over a wide redshift range, where it is necessary to combine different observational datasets \cite[e.g.][]{Behroozi13,Moster13}.
At high redshift, inferred stellar mass functions are typically the only observational constraint available
(as opposed to clustering/weak lensing). For $z>2$, limited depth in rest frame optical bands, as well as complicated selection 
functions, can make measurements of the stellar mass function at low stellar masses very challenging, 
although an encouraging level of consensus has been achieved in recent years \cite[][]{Ilbert13, Muzzin13b, Tomczak14}.

Another way to connect the predicted halo population to the observed stellar population is to build a physical model
that couples dark matter halo merger trees with a simple set of ordinary differential equations that govern the 
exchange of mass, metals and angular momentum between different discrete galaxy and halo components. These
models are typically referred to as semi-analytic galaxy formation models \cite[e.g.][]{Cole00,Somerville08,Guo11}. 
Alternatively, modern computers make it possible to perform hydrodynamical simulations at a resolution capable of resolving galaxies on kpc scales, within
a volume that samples the halo population up to medium size galaxy clusters \cite[][]{Vogelsberger14,Schaye15}. Using either of 
these two modelling techniques, the stellar mass function hosted by a given halo population is predicted and can  
be compared against observational estimates of the stellar mass function without having to assume any parametric 
form for the SHM relation. In general, these modelling techniques have provided support for the parametric forms
assumed in empirical studies \cite[e.g.][]{Zehavi12, Henriques13}.

In this paper, we analyse the predictions made using the semi-analytic model \galform, focussing on
the evolution of the median SHM relation. Unlike other recent work using similar models, we do not attempt to 
find a best-fitting model to some combination of observational data 
\cite[][]{Henriques13,Benson14,Henriques14,Lu14c}. 
Instead, we address the questions: what type of evolution is naturally predicted by semi-analytic 
models for the SHM relation? How much variation in this evolution can be achieved by adjusting model parameters?
What does this evolution tells us about the underlying galaxy formation physics?

Although the model analysed here is just one example of a modern semi-analytic galaxy formation model, 
most of our results can be regarded as fairly general predictions of the semi-analytic
modelling technique (we attempt to point out any obvious exceptions to this at the appropriate points in the text). With that 
said, work from several groups has, in recent years, been focussed on modifying traditional semi-analytic physics parametrisations\footnote{
Which are appropriate for matching the local luminosity/stellar mass function.} for star formation, supernova (SNe) feedback, and gas reincorporation 
in order to try to explain the myriad of observational galaxy evolution results that have been enabled
by recent multi-wavelength surveys \cite[][]{Henriques13,Henriques14,Hirschmann14,Mitchell14,Cousin15b,White15}. While
we do not attempt to explore the breadth of predictions for the evolution of the SHM relation that would result
from exploring all of the modifications that have been suggested (which in some cases are substantial), we do 
present a more limited analysis of the modified gas reincorporation models from our previous work on the star 
forming sequence \cite[][]{Mitchell14}.

The layout of the paper is as follows. In Section~\ref{GALFORM_Section_HM}, we give a brief overview of our reference model. 
In Section~\ref{SHM_section}, we present model predictions for the evolution in the SHM relation. In Section~\ref{simple_explanation}, we
attempt to explain these predictions in simple terms. In Section~\ref{parameters_section}, we assess the impact of changing individual model parameters.
In Section~\ref{AlternativeSection}, we consider the range in SHM evolution that is displayed by a number of models that 
have been roughly tuned to match the local stellar mass function. We discuss and summarise our results in Sections~\ref{discussion}
and \ref{summary} respectively. All data used to produce figures shown in this paper can be made available on request by
contacting the corresponding author (an email address is provided on the first page).

\section{The \galform galaxy formation model}
\label{GALFORM_Section_HM}

In this paper, we explore the predictions for the evolution of the SHM relation made by the semi-analytic galaxy formation model, \galform.
\galform is an example of a model that is built upon the halo merger trees that can be obtained from numerical
simulations or analytical calculations of the hierarchical structure formation that takes place within a $\Lambda$CDM cosmological model. The basis of the model is
that within each subhalo, the baryonic content of galaxies can be compartmentalized into discrete components, including 
disc, bulge and halo components. A set of differential equations
can then be constructed that describe how baryonic mass, angular momentum and metals are exchanged between these discrete components.
The various terms that appear in these continuity equations each represent the effects of a distinct physical process, such as gas cooling or star formation.
A detailed overview of the original implementation of the \galform model can be found in \cite{Cole00}. Significant updates to the physical
modelling are described in \cite{Bower06} (AGN feedback) and
\cite{Lagos11a} (star formation law). An  overview of the most recent
implementation of the model can be found in \cite{Lacey15}. General introductions to semi-analytic 
modelling of galaxy formation can be found in \cite{Baugh06}, \cite{Benson10} and \cite{Somerville14}.

For the reference model used in this paper, we use the model presented in \cite{GonzalezPerez14}.
This model uses merger trees extracted from the {\sc{MR7}}\xspace simulation \cite[][]{Guo13b}, which represents an
update of the {\sc{millennium}}\xspace simulation \citep{Springel05}, using WMAP-7 cosmological parameters \citep{Komatsu11}. As such, unless specified
otherwise, we assume the following cosmological parameters:
$\Omega_{\mathrm{m}}=0.28$, $\Omega_{\mathrm{\Lambda}} = 0.728$, $\Omega_{\mathrm{b}} = 0.045$, $\sigma_{\mathrm{8}} = 0.81$ and $h = 0.704$.

The parameters of the \cite{GonzalezPerez14} model were explicitly tuned to reproduce the observed $b_{\mathrm{J}}$ and $K$-band luminosity 
functions at $z=0$, while also giving reasonable evolution compared to the observed rest-frame UV and $K$-band luminosity functions.
It should be noted that the model was not tuned to reproduce the local stellar mass function inferred from observations.
A comparison between our reference model and observational estimates of the local stellar mass function can be seen in the top-left
panel of Fig.~\ref{smf_model_comp}. Compared to observational estimates, the model underpredicts the abundance of
galaxies at and around the knee of the stellar mass function. Note that in this paper, model predictions for the stellar mass function are always shown using the intrinsic stellar masses from the model (so no attempt is made to replicate the effects of random or systematic measurement error in stellar mass estimates).

For all results presented in this paper, we use corrected DHalo masses to represent the masses of dark matter haloes \cite[][]{Jiang13}.
DHalo masses are defined as the sum of the masses of the subhaloes that the DHalo algorithm associates with a given DHalo. The mass
of each subhalo is defined as the sum of the masses of the particles that are determined to be gravitationally bound to the subhalo by the
{\sc{subfind}}\xspace algorithm \cite[][]{Springel01}\footnote{Note
  that the DHalo mass definition is therefore not equivalent to other commonly used
halo mass definitions such as $M_{200}$ (the mass enclosed within a sphere that has a mean density that is $200$ times the critical density of the Universe).}.
DHalo masses are then corrected (in some cases) to ensure mass conservation such that all haloes grow monotonically in mass in the merger trees.
When quoting halo masses for central galaxies, the halo mass quoted is the corresponding DHalo mass. For satellite galaxies, the halo mass quoted is
the maximum past DHalo mass of the hosting subhalo.
Subhaloes are identified as satellites (for the first time) by the DHalo algorithm if they are enclosed within twice the half-mass radius of a more
massive subhalo and they have lost at least $25 \, \%$ of their past maximum mass \cite[see Appendix A3 in][]{Jiang13}.

The DHalo halo mass definition is similar to but not the same as the conventions followed by the abundance matching studies which we compare 
against later in this paper \cite[][]{Behroozi13,Moster13}. Furthermore, there are significant differences in the abundance of satellite galaxies
between our model and the empirical models of \cite{Behroozi13} and \cite{Moster13}. In Appendix~\ref{ap:msmh}, we present an analysis of this issue,
along with details
of a method to correct for the resultant differences in halo catalogues when comparing predictions for the SHM relation from our model with
the results of abundance matching. When showing results from abundance matching for the evolution of the SHM relation, we show both the
evolution taken directly from \cite{Behroozi13} and \cite{Moster13} and the corresponding evolution we find after applying this correction.

\subsection{Implementation of star formation, SNe feedback and gas reincorporation}
\label{Simple_SFG_explanation_section}

Before presenting our results, it is useful to review the basic
physical processes that regulate the rate ($\dot{M_\star}$) and efficiency ($\dot{M_{\star}}/\dot{M_{\mathrm{H}}}$) of stellar 
mass assembly for actively star forming galaxies in our model. 
For star forming galaxies, where radiative cooling timescales are typically short and AGN feedback is ineffective,
the relevant parts of the model that control the rate and efficiency of stellar mass assembly are the cosmological
infall rate, the star formation law, the efficiency of SNe feedback and the timescale over which the ejected gas is
reincorporated into the gas halo.

Assuming gas traces dark matter accretion rates onto haloes, specific
gas accretion rates, on average, scale strongly with redshift, approximately as $\dot{M_{\mathrm{g}}}/M_{\mathrm{g}} \propto (1+z) H(z)$
in the $\Lambda$CDM model \citep{Fakhouri10}. Once gas is accreted onto a given halo, it takes approximately a single halo dynamical time to 
freefall\footnote{Radiative cooling timescales are almost always shorter than the gravitational freefall time for haloes that host actively star forming galaxies.}
onto the disc at the centre of the halo. The halo dynamical time, $t_{\mathrm{dyn}}$ is defined as

\begin{equation}
t_{\mathrm{dyn}} \equiv \frac{r_{\mathrm{H}}}{V_{\mathrm{H}}} = \frac{G M_{\mathrm{H}}}{V_{\mathrm{H}}^3},
\label{tdyn_eqn}
\end{equation}

\noindent where $r_{\mathrm{H}}$ is the halo virial radius, $V_{\mathrm{H}}$ is the halo circular velocity at that radius,
$M_{\mathrm{H}}$ is the halo mass and $G$ is the gravitational constant.

As introduced in \cite{Lagos11a}, cold gas in galaxy discs is turned into stars at a rate given by the empirical \cite{Blitz06} molecular gas
star formation law,

\begin{equation}
\Sigma_{\mathrm{SFR}} = \nu_{\mathrm{SF}} f_{\mathrm{mol}} \Sigma_{\mathrm{gas}},
\label{blitz_eqn}
\end{equation}

\noindent where $\Sigma_{\mathrm{SFR}}$ is the star formation rate surface density, $\nu_{\mathrm{SF}}$ is the inverse of a characteristic 
star formation timescale, $f_{\mathrm{mol}}$ is the fraction of cold hydrogen gas in the molecular phase and $\Sigma_{\mathrm{gas}}$ 
is the total cold gas surface density. Eqn.~\ref{blitz_eqn} is integrated over the surface of the disc to obtain the star formation rate, $\psi$.
By assuming instantaneous recycling in stellar evolution, the rate of change of stellar mass in the disc is related to $\psi$ by

\begin{equation}
\dot{M}_\star = (1-R) \psi,
\label{mstardot_eqn}
\end{equation}

\noindent where $R$ is the fraction of mass returned to the cold ISM through stellar evolution.

As cold gas forms stars in a disc, a fraction of the cold gas reservoir is continuously ejected from the disc, representing the effects
of SNe feedback. This is quantified by the dimensionless mass loading factor, $\beta_{\mathrm{ml}}$, which is parametrised as a function 
of the disc circular velocity at the half mass radius, $V_{\mathrm{disc}}$, such that

\begin{equation}
\beta_{\mathrm{ml}} = (V_{\mathrm{disc}} / V_{\mathrm{hot}}) ^ {- \alpha_{\mathrm{hot}}},
\label{mass_loading_definition}
\end{equation}

\noindent where $V_{\mathrm{hot}}$ and $\alpha_{\mathrm{hot}}$ are model parameters. The outflow rate from the disc is related to $\beta_{\mathrm{ml}}$
by

\begin{equation}
\dot{M}_{\mathrm{ej}} = \beta_{\mathrm{ml}} \, \psi.
\label{outflow}
\end{equation}

\noindent The effective gas depletion timescale of a galaxy disc, $t_{\mathrm{eff}}$ is therefore given by

\begin{equation}
t_{\mathrm{eff}} = \frac{M_{\mathrm{cold}}}{\psi (1 -R +\beta_{\mathrm{ml}})},
\label{teff_depletion_eqn}
\end{equation}

\noindent where $M_{\mathrm{cold}}$ is the cold gas mass in the disc.

All of the gas that is ejected from a galaxy disc by SNe feedback is then added to a reservoir, $M_{\mathrm{res}}$, of ejected gas which, in turn, is reincorporated back 
into the gas halo at a rate, $\dot{M}_{\mathrm{ret}}$, given by

\begin{equation}
\dot{M}_{\mathrm{ret}} = \frac{\alpha_{\mathrm{reheat}} M_{\mathrm{res}}}{t_{\mathrm{dyn}}},
\label{reincorporation}
\end{equation}

\noindent where $\alpha_{\mathrm{reheat}}$ is a model parameter. In our reference model, $\alpha_{\mathrm{reheat}} = 1.26$,
such that ejected gas is reincorporated back into the halo roughly over a halo dynamical timescale.

To summarise, gas is accreted from the halo onto the disc over roughly a halo dynamical timescale, $t_{\mathrm{dyn}}$. Cold gas is depleted from the disc
over an effective disc depletion timescale, $t_{\mathrm{eff}}$. Given the model parameters, the majority of this cold gas is ejected and subsequently reincorporated into
the halo over roughly a halo dynamical timescale, $t_{\mathrm{dyn}}$. It is important to note that for the haloes hosting star forming galaxies in our reference model, 
SNe feedback is very strong, such that $\beta_{\mathrm{ml}}$ typically significantly exceeds unity. As a result, $t_{\mathrm{eff}}$ tends to be much 
shorter than the other relevant timescale of the system ($t_{\mathrm{dyn}}$). In this regime, the star formation law adopted in 
our model (given by Eqn.~\ref{blitz_eqn}) has minimal impact on the efficiency of stellar mass assembly. Instead, the efficiency is governed by the mass loading factor, 
$\beta_{\mathrm{ml}}$ and the number of times gas can be cycled by feedback through a halo after being accreted 
($\approx t_{\mathrm{H}} / 2 t_{\mathrm{dyn}} \approx 5$). We refer the interested reader to section $4.2$ in \cite{Mitchell14} for a more 
detailed discussion of this point.

\section{The predicted evolution in the SHM relation}
\label{SHM_section}

\begin{figure*}
\begin{center}
\includegraphics[width=40pc]{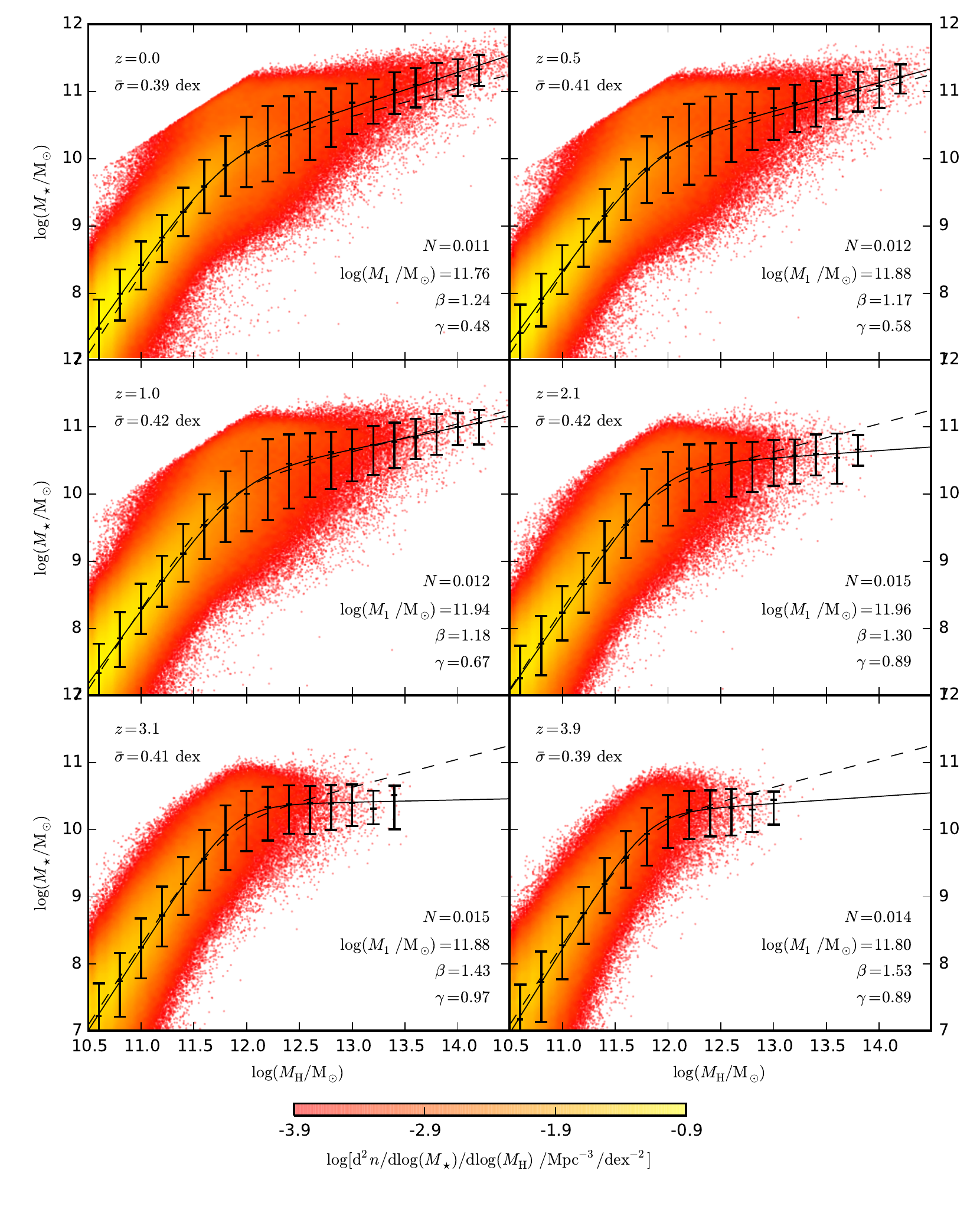}
\caption{Stellar mass plotted as a function of halo mass in our reference model. 
Each panel corresponds to a different redshift, as labelled.
The coloured points represent individual model galaxies and the point colours are scaled with the logarithm of the local point density.
The corresponding number densities are indicated by the colour bar at the bottom of the figure.
The black points and associated error bars show the median, $16^{\mathrm{th}}$ and $84^{\mathrm{th}}$ percentiles of the distribution at a given halo mass.
$\bar{\sigma}$ quantifies the mean scatter in stellar mass within bins of halo mass above $\log(M_{\mathrm{H}}/\mathrm{M_{\odot}})=10.5$.
The scatter in each bin is defined as half of the central $68\%$ range in $\log(M_\star)$.
Black solid lines show the parametrisation given in \protect Eqn.~\ref{shm_eqn}, fit to the medians of the distribution.
The values of $N, M_{\mathrm{1}}, \beta$ and $\gamma$ shown in each panel are the best-fitting parameters from this parametrisation.
Black dashed lines show a similar fit but with the constraint that the fitting parameters do not evolve with redshift.
Each redshift shown is assigned equal weight in the fit. 
The best-fitting parameters for this fit are $N=0.012$, $\log(M_{\mathrm{1}} \, / \mathrm{M_\odot}) = 11.75$, $\beta=1.41$ and $\gamma=0.59$.}
\label{mstar_mhalo_evo}
\end{center}
\end{figure*}

In Fig.~\ref{mstar_mhalo_evo}, we show the evolution in the SHM distribution of our reference model.
For this paper, we are primarily interested in understanding the evolution of the median SHM relation, 
as opposed to the complete SHM distribution. We note however 
that our reference model does not predict that the intrinsic scatter around the median SHM relationship
is strictly lognormal with constant width. This is in contrast to what is assumed in various abundance 
matching studies \cite[][]{Yang12,Behroozi13,Moster13}.
We explore this topic further in Appendix~\ref{ap:shm_distn}.

\begin{figure*}
\begin{center}
\includegraphics[width=40pc]{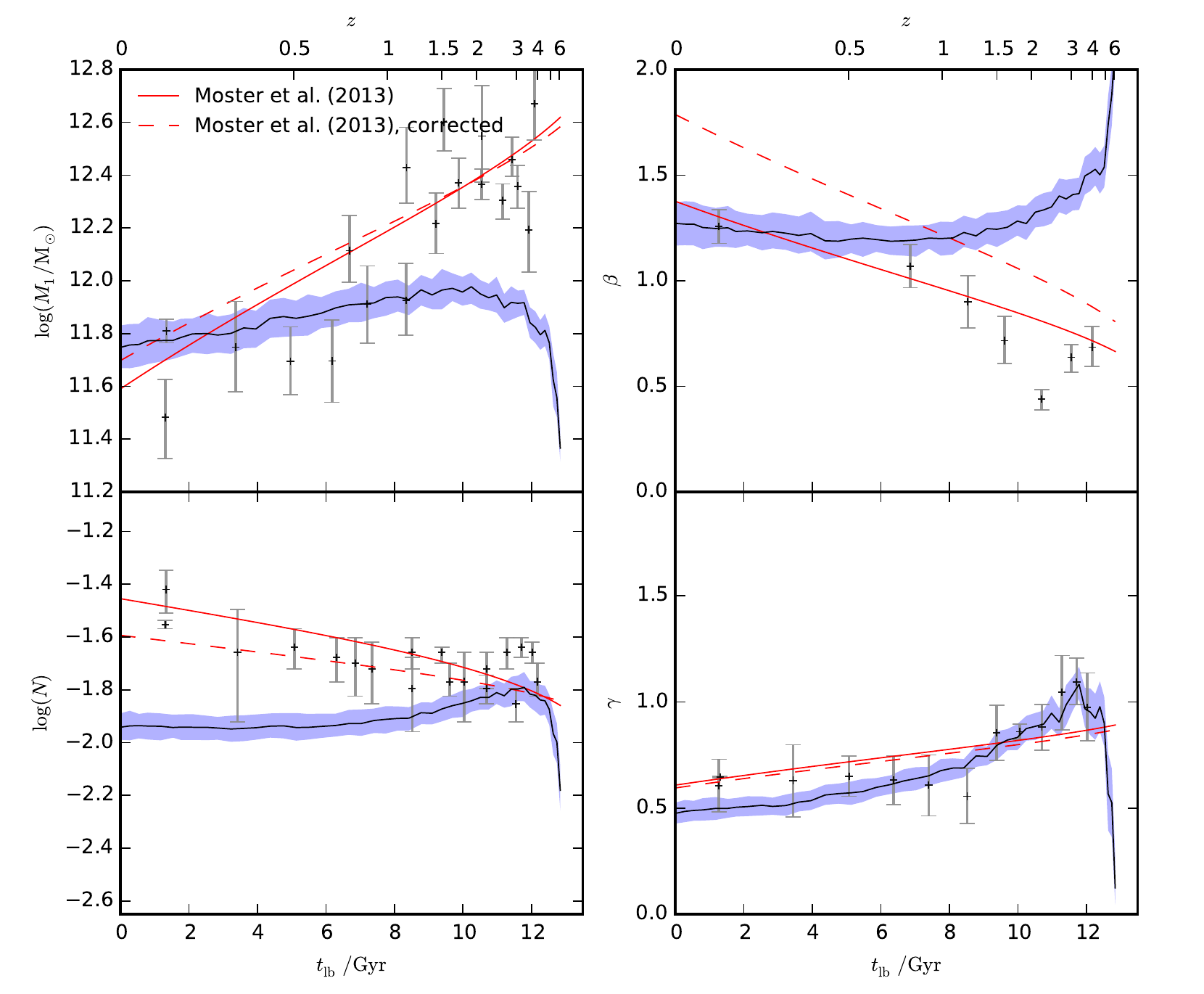}
\caption{Evolution in the fitting parameters for the median relationship between stellar mass and halo mass predicted in our reference model (see \protect Eqn.~\ref{shm_eqn}).
Black solid lines show the median of the projected posterior distribution for each parameter.
Blue shaded regions show the $16^{\mathrm{th}}$ to $84^{\mathrm{th}}$ percentile range of the posterior distributions.
Red solid lines show the best-fitting parametric evolution determined by \protect \cite{Moster13} using multi-epoch abundance matching.
Grey points show the associated best-fitting SHM parameters and $ 1 \sigma$ errorbars determined by \protect \cite{Moster13} using single epoch abundance matching applied to individual stellar mass functions from the literature.
Red dashed lines show the best-fitting parametric evolution we obtain after correcting the \protect \cite{Moster13} SHM relation to be compatible with the halo catalogues used in \galform.}
\label{mstar_mhalo_fit_evo}
\end{center}
\end{figure*}

To quantify the evolution in the median SHM relation, we adopt the parametrisation of \cite{Moster13}, which 
relates the median stellar mass at a given halo mass to halo mass by

\begin{equation}
\frac{M_\star}{M_{\mathrm{H}}} = 2 N \left[ \left(\frac{M_{\mathrm{H}}}{M_{\mathrm{1}}}\right)^{-\beta} + \left(\frac{M_{\mathrm{H}}}{M_{\mathrm{1}}}\right)^{\gamma}  \right]^{-1} ,
\label{shm_eqn}
\end{equation}

\noindent where $N$ is a parameter controlling the normalisation, $M_{\mathrm{1}}$ controls the position of the break\footnote{Note that $M_{\mathrm{1}}$ is closely related to (but not exactly equal to) the characteristic halo mass corresponding to peak stellar mass assembly efficiency (the maximum value of $M_{\mathrm{\star}} / M_{\mathrm{H}}$).}, $\beta$ sets the power law slope below the break and $\gamma$ sets the slope above it.
The meaning of each parameter can be seen in Fig.~\ref{cartoon}.

The evolution in these parameters is shown in Fig.~\ref{mstar_mhalo_fit_evo}, along with abundance matching results 
from \cite{Moster13} for comparison. We show (solid red lines) the evolution in SHM 
parameters using the best-fitting parametric evolution from Table 1 in \cite{Moster13}, which were 
inferred from observational stellar mass function data from \cite{Baldry08}, \cite{PerezGonzalez08}, \cite{Li09} and 
\cite{Santini12}. We also show (dashed red lines) the evolution of SHM fitting parameters which we obtain after correcting for the differences
in input halo catalogues between our reference \galform model and \cite{Moster13}. In effect, this shows the SHM fitting
parameters that \cite{Moster13} would have obtained, had they used our definition of halo mass and our treatment of satellite
galaxies (and their associated subhaloes). The method used to calculate this correction is described in Appendix~\ref{ap:msmh}.

Compared to the results from \cite{Moster13}, our reference model predicts modest evolution
in most of the SHM parameters. In particular, the $1 \, \sigma$ posterior distributions for $\beta$, $M_{\mathrm{1}}$ and $N$ are
consistent with there being no evolution in these parameters for $z < 4$. This is in contrast to the observational abundance matching results,
which suggest comparatively strong evolution in $\beta$ and $M_{\mathrm{1}}$ over the same redshift range.
By comparing solid and dashed red lines, it appears that this difference between our model and abundance matching is robust against differences in the input halo catalogues.
This demonstrates that there are differences between our reference model and the \cite{Moster13} empirical model that are caused instead
by the details of the implementation of baryonic physics in our reference model.

\begin{figure*}
\begin{center}
\includegraphics[width=40pc]{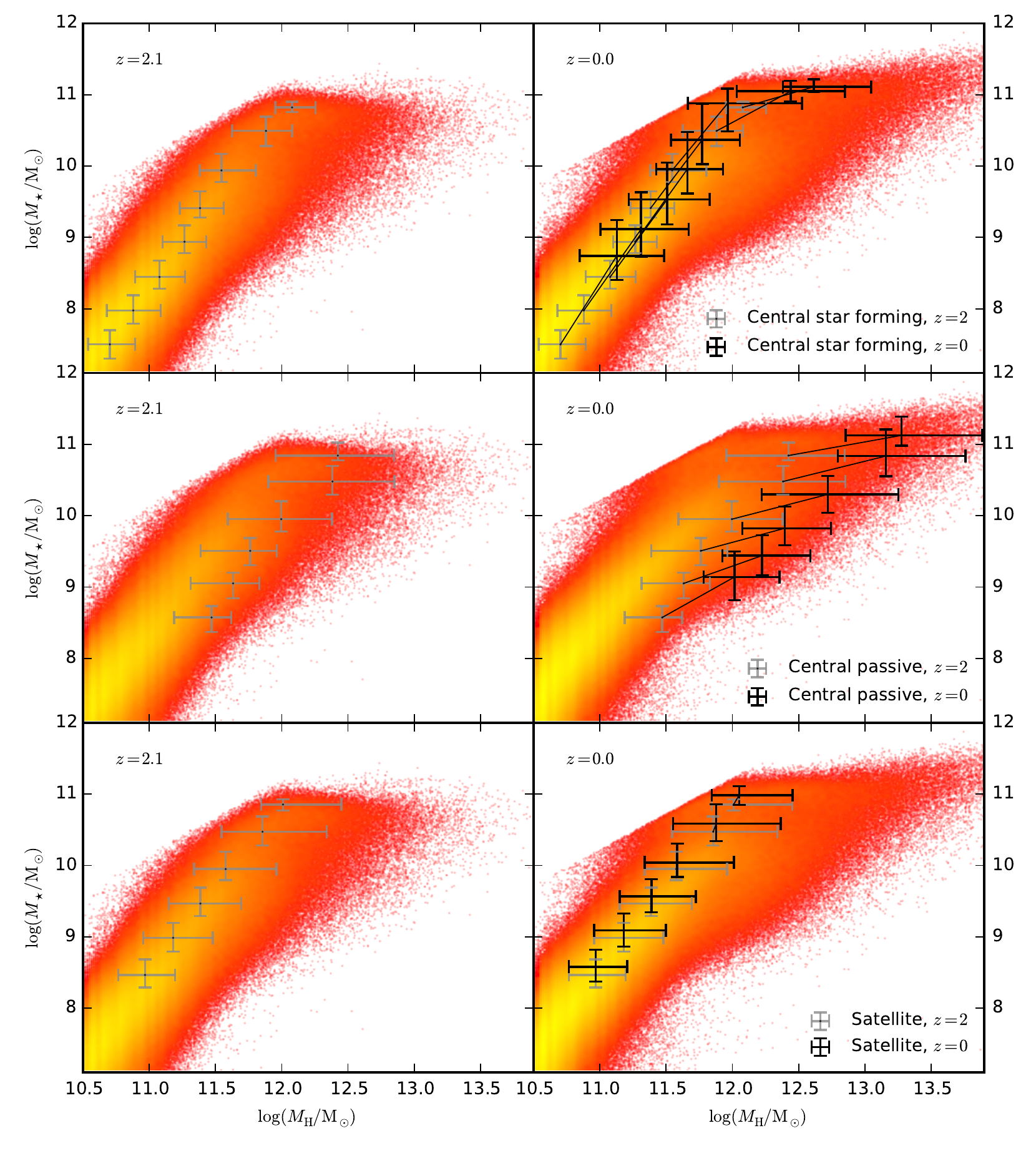}
\caption{Evolution of populations of galaxies across the stellar mass versus halo mass plane in our reference model.
In all panels, the coloured points show the distribution for the entire galaxy population at the redshift labelled at the top-left corner of each panel.
Subpopulations of galaxies are selected in stellar mass bins at $z=2$ (grey error bars). Each subpopulation is then tracked to $z=0$ (black error bars) and
the intervening evolution of each subpopulation is then indicated by solid black lines connecting grey and black error bars.
{\it Top}: Evolution of central star forming galaxies. These are galaxies that are star forming for at least $90 \%$ of the simulation outputs between $z=2$ and $z=0$.
{\it Middle}: Evolution of central passive galaxies. These are central galaxies that are star forming before $z \sim 2$ but are then passive after $z \sim 2$.
{\it Bottom}: Evolution of satellite galaxies. These are galaxies that are central before $z \sim 2$ that then become satellites after $z \sim 2$. 
The error bars show the $10^{\mathrm{th}}$ and $90^{\mathrm{th}}$ percentiles in both stellar mass and halo mass for each subpopulation and stellar mass bin.}
\label{mstar_mhalo_combined}
\end{center}
\end{figure*}

To try to understand the reasons for the modest evolution in the SHM relation predicted by our 
reference model, we split the overall population into subsamples of central star forming, central passive and satellite
galaxies. To split star forming from passive galaxies, we use the same evolving cut in specific star formation rate against
stellar mass that was used in \cite{Mitchell14}\footnote{The analytic evolution of the cut is designed by hand to
separate the distributions of star forming and passive galaxies at all redshifts considered.}.
The evolution in the SHM relations for these subsamples is shown in Fig.~\ref{mstar_mhalo_combined}. 
Starting with central star forming galaxies, in the top panels of 
Fig.~\ref{mstar_mhalo_combined} we show how a sample of model star forming galaxies has evolved since $z=2$. 
Specifically, we select central galaxies that are classed as star forming for at least $90 \%$ of the output times in our 
merger trees between $z=2$ and $z=0$.
Fig.~\ref{mstar_mhalo_combined} shows directly that these model galaxies essentially evolve 
along an invariant power law in the $M_\star$ versus $M_{\mathrm{H}}$ plane. This power law is consistent with the overall SHM
relation below the break mass ($M_{\mathrm{1}}$), explaining why the overall SHM relation does not evolve significantly
in this halo mass range. We note that this phenomenon has also been inferred from observational weak lensing data between $z=1$ and $z=0$ by \cite{Hudson15}.
This behaviour is, however, broken in the two highest stellar mass bins (spanning $10.25 < \log(M_\star \, \mathrm{M_\odot}) < 11.25$)
of central star forming galaxies selected at $z=2$. Here, the fractional growth in stellar mass is outpaced by the fractional growth of dark matter haloes.

In the middle panels of Fig.~\ref{mstar_mhalo_combined}, we show a sample of central model galaxies that are star forming before $z=2$ and then 
become passive at $z<2$. Specifically, we select galaxies that are star forming for $> 90 \%$ of the simulation output
times for $z>2$ and are passive for $> 90 \%$ of the simulation output times for $z<2$.
By $z=0$, these galaxies are displaced from the median of the SHM distribution for star forming galaxies, preferentially
residing in more massive haloes at a given stellar mass. It is apparent that these passive galaxies do not follow the same evolutionary
path of all but the most massive star forming galaxies shown in Fig.~\ref{mstar_mhalo_combined}. Instead, the growth in their host dark matter haloes
outpaces any stellar mass assembly through galaxy mergers. This behaviour helps to creates the break in the overall SHM relation above $M_{\mathrm{H}} = M_{\mathrm{1}}$, 
where passive central galaxies dominate the overall population.

Finally, for completeness, in the bottom panels of Fig.~\ref{mstar_mhalo_combined} we show a sample of model galaxies that are central before $z=2$
and then become satellites after $z=2$. Specifically, we select galaxies that are central for $> 90 \%$ of the simulation output
times before $z=2$ and are satellites for $> 90 \%$ of the simulation output times for $z<2$. Fig.~\ref{mstar_mhalo_combined}
shows the expected result that our model predicts that satellite galaxies do not grow significantly in stellar mass after infall. 
This result is expected because of the implementation of hot gas stripping and SNe feedback in this version of \galform
\footnote{Hot gas is instantaneously stripped from satellite haloes and strong SNe feedback typically ejects the majority of the cold gas on a short timescale.}.
By definition, for the SHM relation, satellite halo masses are set as the mass of the associated subhalo at infall. Without any significant star formation 
activity after infall, satellites therefore simply remain frozen in place in the SHM plane. As the SHM 
relation below the break does not evolve significantly in our reference model, satellites do not become significantly displaced from the
total SHM distribution after infall. 

We note that the instantaneous hot gas stripping used in our reference model is unlikely to be realistic \citep{Font08,Henriques14}. 
Observational data suggest that satellite galaxies typically continue to form stars at a comparable rate to central galaxies for a significant
length of time after infall \cite[e.g.][]{Peng10,Wetzel12,Wetzel13,McGee14}. However, based on results from \cite{Watson13}, who use galaxy clustering and
group catalogues to show that the SHM relation for satellite galaxies is consistent with the SHM relation for central galaxies over $0<z<2$, we do 
not expect that the abrupt quenching of satellite galaxies after infall in our model will adversely affect predictions for the median SHM relation.

\section{Physical reasons for the lack of evolution in the predicted SHM relation}
\label{simple_explanation}

In Section~\ref{SHM_section}, we showed that our reference model predicts that, below a break halo mass, $M_{\mathrm{1}}$, the median 
SHM relationship does not evolve significantly over $0<z<4$. In this section, we attempt to explain, in simple terms, why our reference 
model predicts minimal evolution in the SHM relation below $M_{\mathrm{1}}$, the origin of the break mass, $M_{\mathrm{1}}$, and why the 
predicted high mass slope of the SHM relation above the break increases with cosmic time. To do so, we consider the impact of SNe
feedback, AGN feedback and galaxy mergers on the SHM relation predicted by our reference model. For reasons of clarity, we introduce
(for this section only) two new variables, $\beta' \equiv 1+\beta$ and $\gamma' \equiv 1 - \gamma$ (where $\beta$ and $\gamma$ are
parameters from the fitting formula given by Eqn.~\ref{shm_eqn}). $\beta'$ and $\gamma'$ are the power slopes of the SHM relation such that 
$M_\star \propto M_{\mathrm{H}}^{\beta'}$ for $M_{\mathrm{H}} \ll M_{\mathrm{1}}$ and $M_\star \propto M_{\mathrm{H}}^{\gamma'}$ for $M_{\mathrm{H}} \gg M_{\mathrm{1}}$.

\subsection{Star forming galaxies}
\label{sfg_msmh_invariance_section}

Below a break halo mass, $M_{\mathrm{1}}$, Fig.~\ref{mstar_mhalo_fit_evo} shows that our reference model does not predict significant evolution in the SHM relation
over $0<z<4$. For halo masses below $M_{\mathrm{1}}$, the galaxy population in our model is dominated by central star forming
and satellite galaxies (as opposed to central passive galaxies). In this halo mass regime, the median SHM relation is well
described by a non-evolving power law with slope $\beta' = 1+\beta \approx 2.3$ for  $z < 4$. This is interesting, given that for star 
forming galaxies, it might be expected in the simplest possible case that the star formation rate, $\dot{M}_\star$, would simply 
track the accretion rate onto haloes, $\dot{M}_\mathrm{H}$. In this case, individual galaxies would evolve 
along a power law in the SHM plane with slope, $\beta' = 1$. To evolve along a power law where $\beta' \approx 2.3$ 
requires instead that 

\begin{equation}
\dot{M}_\star \propto M_{\mathrm{H}}^{1.3} \dot{M}_\mathrm{H}, 
\label{shm_evo_eqn}
\end{equation}

\noindent implying that stellar mass assembly increases in efficiency as the host haloes grow in mass.
It should be noted that this also requires that at fixed halo mass, the instantaneous star formation efficiency,

\begin{equation}
\eta_{\mathrm{SF}} \equiv \dot{M}_\star / (f_{\mathrm{B}} \dot{M}_{\mathrm{H}}),
\label{sf_efficiency_msmh}
\end{equation}

\noindent is constant across cosmic time (here, $f_{\mathrm{B}}$ is the cosmic baryon fraction).
In \cite{Mitchell14}, we showed that $\eta_{\mathrm{SF}}$ does evolve for populations 
of star forming galaxies in \galform as their haloes grow in mass. This occurs predominantly because of 
evolution in the mass loading factor for SNe feedback\footnote{We note that for some of the discussion 
presented in \cite{Mitchell14}, we further simplified this picture by assuming that $\beta_{\mathrm{ml}}$ 
is constant over some redshift range, which results in $\dot{M}_\star/M_\star \propto \dot{M}_{\mathrm{H}} / M_{\mathrm{H}}$.
While this is approximately true at lower redshifts, it is an oversimplification when considering 
the evolution of the SHM relation.}, $\beta_{\mathrm{ml}}$, although a small amount of evolution in the gas reincorporation 
efficiency also contributes \cite[][]{Mitchell14}.

In \galform, $\beta_{\mathrm{ml}} \propto V_{\mathrm{disc}}^{-\alpha_{\mathrm{hot}}}$, where $V_{\mathrm{disc}}$ is the disc
circular velocity at the half-mass radius. Roughly speaking, the disc circular velocity scales with the halo circular velocity, $V_{\mathrm{H}}$, in 
smaller haloes where baryon self gravity effects are not important. These are the haloes that typically host star 
forming galaxies and also the haloes where SNe feedback plays the largest role in regulating star formation 
rates. Again, roughly speaking\footnote{This is not a precise statement because finite gas reincorporation 
and freefall/radiative cooling timescales will cause the full effects of any instantaneous changes 
in $\beta_{\mathrm{ml}}$ to take time to propagate through the system of equations.}, it is expected that 
the instantaneous star formation efficiency, $\eta_{\mathrm{SF}}$, will scale with $(1+\beta_{\mathrm{ml}})^{-1} \approx \beta_{\mathrm{ml}}^{-1}$
\cite[][]{Mitchell14}. 
Therefore, it is to be expected that in low mass haloes that host star forming galaxies, 

\begin{equation}
\eta_{\mathrm{SF}} \propto \beta_{\mathrm{ml}}^{-1} \propto V_{\mathrm{disc}}^{\alpha_{\mathrm{hot}}} \propto V_{\mathrm{H}}^{\alpha_{\mathrm{hot}}} \propto M_{\mathrm{H}}^{\alpha_{\mathrm{hot}} /3} \, \bar{\rho}_{\mathrm{H}}^{\alpha_{\mathrm{hot}} /6},
\label{simple_scaling}
\end{equation}

\noindent where $\bar{\rho}_{\mathrm{H}}$ is the mean halo density, which is related to $V_{\mathrm{H}}$ through

\begin{equation}
\bar{\rho}_{\mathrm{H}} = \frac{3 M_{\mathrm{H}}}{4 \pi R_{\mathrm{H}}^3}, \quad  V_{\mathrm{H}}^2 = \frac{G M_{\mathrm{H}}}{R_{\mathrm{H}}}.
\label{rho_halo}
\end{equation}

\noindent $\bar{\rho}_{\mathrm{H}}$ is independent of halo mass, but does instead depend on the expansion factor, $a$, through

\begin{equation}
\bar{\rho}_{\mathrm{H}} = \Delta_{\mathrm{vir}}(a) \rho_{\mathrm{crit}}(a),
\label{rho_halo2}
\end{equation}

\noindent where $\Delta_{\mathrm{vir}}(a)$ is the overdensity of collapsed haloes relative to the critical density and $\rho_{\mathrm{crit}}(a)$ is the critical density
of the Universe.

If we make the approximation that the rate of halo mass growth outpaces halo density growth,
such that $\bar{\rho}_{\mathrm{H}}$ is effectively constant over the timescale for haloes to 
assemble their mass, then straightforward integration
of Eqn.~\ref{simple_scaling} yields 

\begin{equation}
M_\star \propto M_{\mathrm{H}}^{1 + \alpha_{\mathrm{hot}}/3},
\label{non_evolv_shm_eqn}
\end{equation}

\noindent which, for the value of $\alpha_{\mathrm{hot}} = 3.2$ used in our reference model, yields 
$M_\star \propto M_{\mathrm{H}}^{2.07}$. This implies $\beta'=2.07$, close to the value, $\beta'=2.3$ 
predicted by our reference model. Therefore, we see that the slope of the SHM relation primarily reflects
the exponent, $\alpha_{\mathrm{hot}}$, in the SNe feedback mass loading parametrisation. 
§
We note that the mean halo density, $\bar{\rho}_{\mathrm{H}}$, is not constant across cosmic time. However, in 
Appendix~\ref{ap:msmh2} we explain why this is not a bad approximation when integrating Eqn.~\ref{simple_scaling}.
We show that for $z>1$, halo mass accretion rates greatly outpace the rate of change in $\bar{\rho}_{\mathrm{H}}$.
Another important effect is that, on average, halo densities evolve more slowly for haloes in \galform
when compared to the spherical collapse model. This is because halo circular velocities (and hence mean halo 
densities at fixed halo mass) in our reference model are only updated to match the spherical collapse model when haloes 
double in mass. This effect is particularly important for $z<1$ when halo mass doubling events are infrequent (as
halo mass accretion rates have decreased compared to high redshift).

\subsection{AGN feedback}
\label{agn_fb_section}

We now consider the origin of the break halo mass, $M_{\mathrm{1}}$, which marks the point above which
the efficiency of stellar mass assembly drops with respect to the efficiency of halo mass assembly.
Moreover, we seek to explain why $M_{\mathrm{1}}$ does not evolve significantly in our model for $z<4$ (see 
Fig.~\ref{mstar_mhalo_fit_evo}). 

In our model, the efficiency of stellar mass assembly drops in massive, quasi-hydrostatic haloes 
because AGN feedback acts to shut down cooling from hot gas coronae onto the central galaxy. For the 
AGN feedback model implemented in \galform, the primary requirement for AGN feedback to be effective
in suppressing cooling is that a halo is in quasi-hydrostatic equilibrium, which is taken to be true
if

\begin{equation}
t_{\mathrm{cool}}(r_{\mathrm{cool}}) > \alpha_{\mathrm{cool}}^{-1} \, t_{\mathrm{ff}}(r_{\mathrm{cool}}),
\label{agn_fb_eqn}
\end{equation}

\noindent where $t_{\mathrm{cool}}$ is the radiative cooling timescale evaluated at a radius $r_{\mathrm{cool}}$, 
$\alpha_{\mathrm{cool}}$ is a model parameter and $t_{\mathrm{ff}}$ is the gravitational freefall timescale evaluated at $r_{\mathrm{cool}}$ in a 
NFW halo. The cooling radius $r_{\mathrm{cool}}$ is defined as the radius below which the enclosed gas had sufficient
time to cool through radiative processes since the previous halo formation event \cite[][]{Cole00}.

On average, Eqn.~\ref{agn_fb_eqn} sets a threshold halo mass above which AGN feedback
is effective. A simplified expectation for how this AGN feedback threshold mass evolves with time can be obtained
by evaluating $t_{\mathrm{cool}}$ at the mean gas density within the halo, $\bar{\rho}_{\mathrm{g}}$.

\begin{equation}
t_{\mathrm{cool}}(\rho_{\mathrm{g}} = \bar{\rho}_{\mathrm{g}}) = \frac{3}{2} \frac{k_{\mathrm{B}} T}{\mu m_{\mathrm{p}}} \frac{1}{\bar{\rho}_{\mathrm{g}} \Lambda(Z_{\mathrm{g}},T)},
\label{cooling_time_chHM}
\end{equation}

\noindent where $\Lambda(Z_{\mathrm{g}},T)$ is the cooling function, which depends on the hot gas metallicity, $Z_{\mathrm{g}}$,
and temperature, $T$, evaluated at the mean gas density. Assuming the gas temperature is equal to the virial temperature
of the halo, $T_{\mathrm{vir}}$, given by

\begin{equation}
T_{\mathrm{vir}} = \frac{1}{2} \frac{\mu m_{\mathrm{p}}}{k_{\mathrm{B}}} V_{\mathrm{H}}^2,
\label{T_vir}
\end{equation}

\noindent we obtain the scaling that

\begin{equation}
t_{\mathrm{cool}} \propto \frac{V_{\mathrm{H}}^2}{\bar{\rho}_{\mathrm{g}} \Lambda(T,Z)} \propto \frac{V_{\mathrm{H}}^2}{\bar{\rho}_{\mathrm{H}} \Lambda(T,Z)} \propto \frac{M_{\mathrm{H}}^{2/3}}{\bar{\rho}_{\mathrm{H}}^{2/3}} \frac{1}{\Lambda(T,Z)}.
\label{tcool_scale}
\end{equation}

\noindent For a fixed NFW halo concentration, the freefall time scales with the halo dynamical 
timescale, $t_{\mathrm{dyn}} = G M_{\mathrm{H}} / V_{\mathrm{H}}^3$, such that

\begin{equation}
t_{\mathrm{ff}} \propto t_{\mathrm{dyn}} \propto \frac{M_{\mathrm{H}}}{V_{\mathrm{H}}^3} \propto \bar{\rho}_{\mathrm{H}}^{-1/2}.
\label{t_ff_scaling}
\end{equation}

\noindent We can then evaluate Eqn.~\ref{agn_fb_eqn} for $t_{\mathrm{cool}} = t_{\mathrm{ff}}$, yielding
a threshold halo mass for effective AGN feedback that evolves according to 

\begin{equation}
\frac{M_{\mathrm{H}}^{2/3}}{\bar{\rho}_{\mathrm{H}}^{2/3}} \frac{1}{\Lambda(T,Z)} \propto \bar{\rho}_{\mathrm{H}}^{-1/2},
\label{agn_scaling_1}
\end{equation}

\noindent which implies that

\begin{equation}
M_{\mathrm{H}} \propto \Lambda(T,Z)^{3/2} \bar{\rho}_{\mathrm{H}}^{1/4} \propto \Lambda(T,Z)^{3/2} \left[ \Delta_{\mathrm{c}}(a) \rho_{\mathrm{crit}}(a) \right]^{1/4}.
\label{agn_scaling_2}
\end{equation}

\noindent In other words, we expect AGN feedback to suppress cooling (and therefore star formation) above a
characteristic halo mass which is only weakly dependent on redshift ($M_{\mathrm{H}} \propto \left[\Delta_{\mathrm{c}}(a) \bar{\rho}(a) \right]^{1/4}$),
for a fixed hot gas metallicity and temperature.
This simple expectation is qualitatively consistent with the modest evolution in the SHM break mass predicted in our reference model.
A more detailed exploration of the behaviour of AGN feedback in our model is presented in Appendix~\ref{ap:msmh4}.
We also refer the interested reader to the discussion presented in section $6.6$ of \cite{Leauthaud12} on whether
the efficiency of stellar mass assembly efficiency actually peaks at a given $M_\star / M_{\mathrm{H}}$ ratio,
rather than at fixed halo mass, as we have described here.

\subsection{Mergers}

Finally, we give brief consideration to the evolution of the high mass SHM slope, $\gamma'$, which is predicted to evolve 
from $\gamma' \approx 0$ at $z=4$ to $\gamma' \approx 0.5$ at $z=0$ in our reference model. In this halo mass regime above 
the SHM break, effective AGN feedback means that stellar mass assembly is 
dominated by galaxy mergers rather than by star formation. Without galaxy mergers, stellar mass assembly would 
stop entirely, such that passive galaxies would evolve along a power law with exponent $\gamma' = 0$ as their host haloes continue 
to grow in mass. In the opposite extreme where infalling satellite galaxies instantly merge onto the central galaxy after a halo merger, the expectation is instead
that stellar mass assembly will simple trace the hierarchical halo assembly process. In this case, passive central galaxies evolve along a power
law with exponent $\gamma' = 1$. 

In reality, satellite galaxies merge a finite period of time after infall. The evolution in our model
form $\gamma' \approx 0$ at $z=4$ to $\gamma' \approx 0.5$ at $z=0$ therefore simply reflects that there
is a latency between halo and satellite galaxy mergers, such that $\gamma'$ is pushed to higher values with
cosmic time.

\section{Dependence on individual model parameters}
\label{parameters_section}

The top panels in Fig.~\ref{mstar_mhalo_combined} shows that in our reference model, the lack of significant evolution in the predicted median SHM relation 
is driven primarily by a characteristic evolutionary path that star forming galaxies follow across the SHM plane. For star 
forming galaxies to evolve in this way requires a fairly specific evolution in the instantaneous
star formation efficiency, $\eta_{\mathrm{SF}}$. This raises the question of whether this characteristic evolutionary path
is a general prediction made by semi-analytic galaxy formation models such as \galform, or just a feature specific to the combination of model parameters used in our reference model.

To address this question, we now explore the evolution of the SHM relation predicted by models with alternative sets of
model parameters. Changing individual model parameters in isolation will typically result in models that give a poor 
match to the local galaxy luminosity function. Nonetheless, this exercise is still useful for giving us an idea about the effect 
that each parameter has on the evolution of the median SHM relationship. A list of the model parameters which we consider for 
this exercise is presented in Table~\ref{model_param_table}.

\begin{table*}
\begin{center}
\begin{tabular}{c c p{10cm}}
  \hline
  $V_{\mathrm{hot}}$ & $133, 425, 600 \, \mathrm{km \, s^{-1}}$ &    Normalisation of SNe feedback. Change such that the mass loading factor, $\beta_{\mathrm{ml}}$, changes up or down by a factor of three.\\
  \hline
  $\alpha_{\mathrm{hot}}$ & $1.6, 3.2, 4.8$ & Dependence of SNe feedback on galaxy circular velocity. Change up or down by $\pm 50 \%$. 
  Also change $V_{\mathrm{hot}}$ such that $\beta_{\mathrm{ml}}$ is fixed for a circular velocity, $V_{\mathrm{disc}} = 200 \, \mathrm{km \, s^{-1}}$.\\
  \hline
  $\alpha_{\mathrm{reheat}}$ & $0.42, 1.26, 3.78$ & Ejected gas reincorporation rate. 
  Change such that $1 + 2 \pi \alpha_{\mathrm{reheat}}$ changes up or down by a factor of three.
  This factor corresponds to the approximate reincorporation efficiency, given by \protect{equation~(22) in \cite{Mitchell14}}.\\
  \hline
  $\nu_{\mathrm{sf}}$ & $0.17, 0.5, 1.7 \, \mathrm{Gyr^{-1}}$ & Disc SF law normalisation.\\
  \hline
  $\alpha_{\mathrm{cool}}$ & $0.2, 0.6, 1.8$ & AGN feedback cooling suppression threshold.\\
  \hline
  $\eta_{\mathrm{disc}}$ & $0.61, 0.8, 2.4$ & Disc instability threshold. Change up by a factor of three and down to 0.61 (minimum value below which all discs are stable).\\
  \hline
  $f_{\mathrm{dyn}}$ & $3.3, 10, 30$ & Burst duration factor.\\
  \hline
  $f_{\mathrm{df}}$ & $0.5, 1.5, 4.5$ & Rescaling factor for the dynamical friction timescale.\\
  \hline
\end{tabular}
\end{center}
\caption{Description of the model parameters that are varied in \protect Section~\ref{parameters_section} to produce the set of models shown in \protect Fig.~\ref{mstar_mhalo_fit_model_comp} and Fig.~\ref{mstar_mhalo_bin_model_comp}.
In all cases, the reference model is the intermediate model for the quoted parameter values.}
\label{model_param_table}
\end{table*}

\begin{figure*}
\begin{center}
\includegraphics[width=40pc]{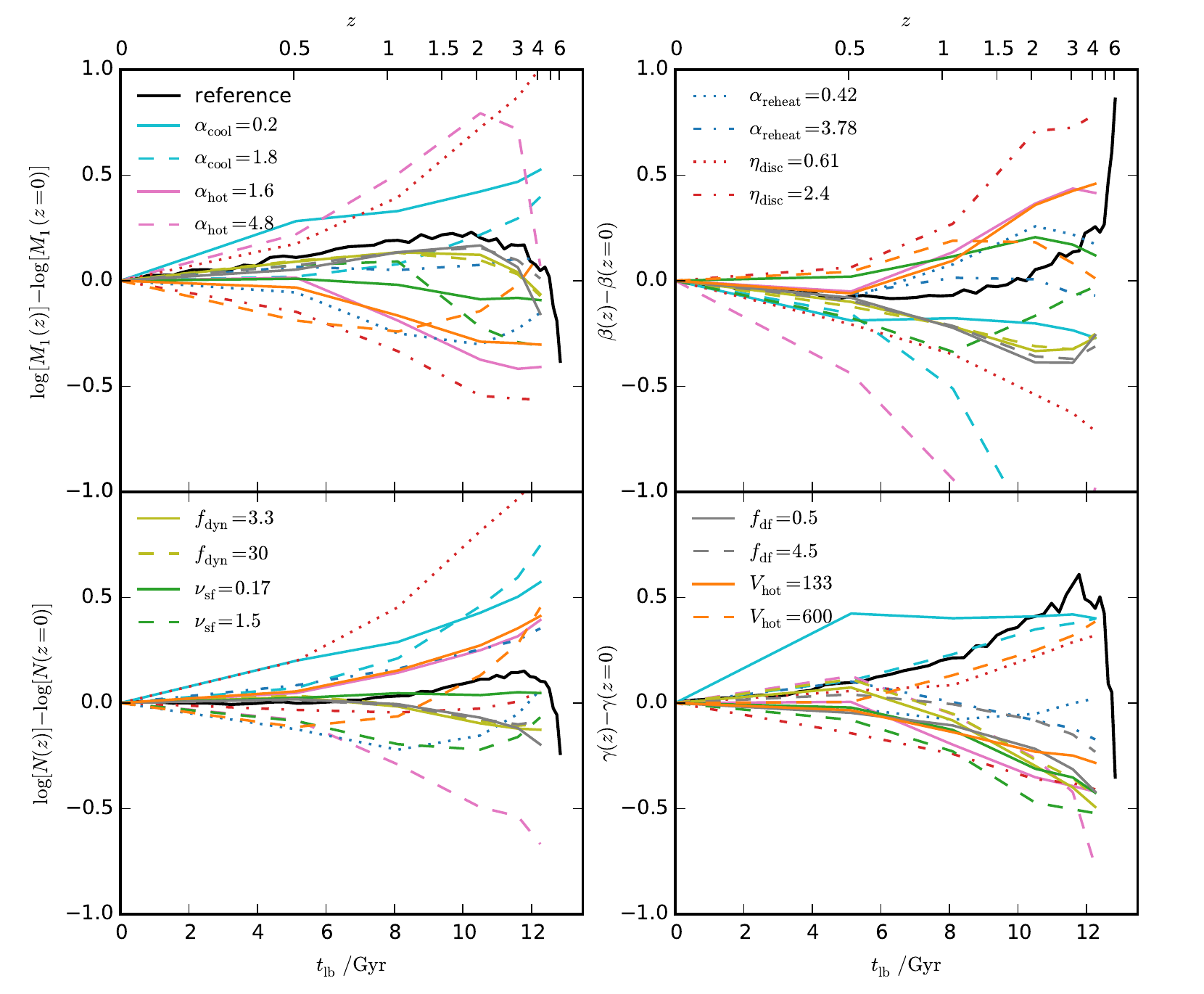}
\caption{Evolution with respect to $z=0$ of fitting parameters for the median SHM relation (see \protect Eqn.~\ref{shm_eqn}).
Each line shows the median of the projected posterior distribution for a given parameter and model, as labelled.
Each model has a single parameter varied with respect to the reference model (black line), as described in \protect Table~\ref{model_param_table}.
Note that the key is spread over all four panels.}
\label{mstar_mhalo_fit_model_comp}
\end{center}
\end{figure*}

\begin{figure*}
\begin{center}
\includegraphics[width=40pc]{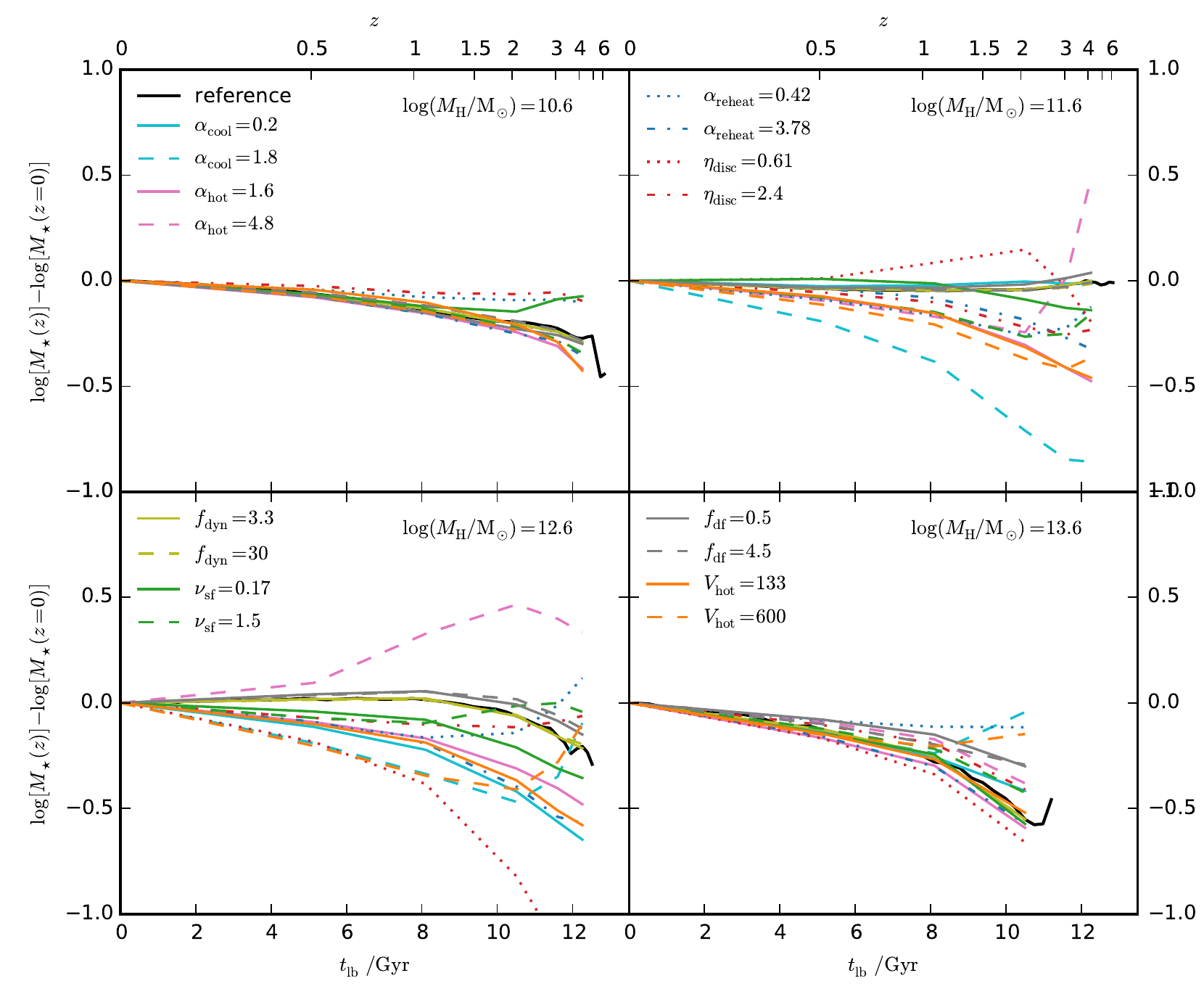}
\caption{Evolution with respect to $z=0$ in the median stellar mass within a given halo mass bin.
Each panel corresponds to a different halo mass bin, as labelled.
Each line shows the evolution in the median stellar mass for a given model, relative to the median stellar mass at $z=0$.
Each model has a single parameter varied with respect to the reference model (black line), as described in \protect Table~\ref{model_param_table}.
Note that the key is spread over all four panels.}
\label{mstar_mhalo_bin_model_comp}
\end{center}
\end{figure*}

The results for the range in evolution of the SHM relation predicted by this model suite are shown in 
Fig.~\ref{mstar_mhalo_fit_model_comp} and Fig.~\ref{mstar_mhalo_bin_model_comp}. 
In Fig.~\ref{mstar_mhalo_fit_model_comp}, we show the evolution in the fitting parameters for 
the parametric SHM relation given by Eqn.~\ref{shm_eqn}. Comparing each variant model in turn with
the reference model, it is clear that the modest evolution in the SHM relation predicted by the
reference model is not a general prediction of \galform. Instead, the reference model appears to 
occupy a unique position in the overall parameter space (at least for a subset of the model parameters).
It should be noted that the parameter variations that we consider here are large. In our
  experience, these large parameter variations can push the model outside of the regime of efficiencies and timescales
  occupied by our reference model (that lead to a non-evolving
  SHM relation). This is the regime of short gas depletion timescales in galaxy discs, strong SNe feedback that scales exactly
  with halo properties, constant gas reincorporation efficiency, and where AGN feedback is very efficient
  when haloes become hydrostatic.

Fig.~\ref{mstar_mhalo_bin_model_comp} shows an alternative view of the range in evolution seen in
Fig.~\ref{mstar_mhalo_fit_model_comp},
this time considering the evolution in fixed halo mass bins. Here, it becomes apparent that our 
reference model is most distinct from at least some of the variant models presented in Table~\ref{model_param_table}
in the $\log(M_{\mathrm{H}})=11.6, \, 12.6$ mass range. We note that these are the bins that approximately 
bracket the SHM break mass, $M_{\mathrm{1}}$. For the other two bins at the lowest and highest halo masses
considered ($\log(M_{\mathrm{H}})=10.6, \, 13.6$), our reference model is more typical of the variant
models we consider here for the predicted evolution.

\section{Alternative models}
\label{AlternativeSection}

Fig.~\ref{mstar_mhalo_fit_model_comp} and Fig.~\ref{mstar_mhalo_bin_model_comp} show that the
small amount of evolution in the SHM relation seen for our reference model is not a general 
prediction of all \galform models. However, the set of models considered in 
Section~\ref{parameters_section} did not, in general, produce an acceptable match to the local 
stellar mass function. This then raises the question of how much variation in the evolution of the 
SHM relation can be predicted by a family of models that do provide an adequate fit to the local 
stellar mass function inferred from observations. Another way to phrase this question is to ask the following:
to what extent does the form of the local stellar mass function inferred from observations constrain 
galaxy formation models to predict a specific type of evolution in the SHM relation?

To answer this question properly would require constructing a full posterior distribution from
the model parameter space to find all acceptable models, using the local stellar 
mass function as a constraint. Here, we take an intermediate step by instead considering only a limited number 
of different models which have been tuned to roughly match the local stellar mass function. 
These models encapsulate some of the variations which, through experience, we expect to be interesting within 
the context of exploring why our reference model predicts very little evolution in the SHM relation. 

\begin{table}
\begin{tabular}{ccccccc}
  \hline
  Model Parameter & Reference & SFB & WFB & M14 & SFH & VM  \\
  \hline
  $\alpha_{\mathrm{hot}}$ & 3.2 & 3.2 & 3.2 & 3.2 & 3.2 & 3.2 \\
  \hline
  $V_{\mathrm{hot}} \, / \mathrm{km \, s^{-1}}$ & 425 & 700 & 300 & 485 & 485 & 485 \\
  \hline
  $\alpha_{\mathrm{reheat}}$ & 1.26 & 8.0 & 0.3 & 1.26 & 0.023 & 0.24 \\
  \hline
  $\alpha_{\mathrm{cool}}$ & 0.6 & 0.65 & 0.4 & 1.0 & 1.3 & 1.0 \\
  \hline
\end{tabular}
\caption{Model parameters used in the five variant models explored in \protect Section~\ref{AlternativeSection}.
$\alpha_{\mathrm{hot}}$ sets the mass loading dependence on circular velocity, $V_{\mathrm{hot}}$ sets the mass loading
normalisation, $\alpha_{\mathrm{reheat}}$ sets the reincorporation rate and $\alpha_{\mathrm{cool}}$ controls the AGN
feedback threshold.
The variant models considered here include the strong and weak feedback models (SFB and WFB), which feature stronger/weaker feedback but with shorter/longer reincorporation timescales to compensate. 
M14 is the reference model from \protect \cite{Mitchell14}.
The star formation history (SFH) and virial mass (VM) models feature different modifications to the reincorporation timescale, as described in the text.}
\label{ModelsTable}
\end{table}

Specifically, we consider five additional models, with model parameters outlined 
in Table~\ref{ModelsTable}. Two of these models represent variations of the reference model. They use 
the same physics parametrisations as the reference model. These two variant models are 
chosen to highlight that there is a degeneracy between the reincorporation rate coefficient, 
$\alpha_{\mathrm{reheat}}$, and the normalisation of the mass loading factor, $V_{\mathrm{hot}}$. 
By either raising or lowering both of these parameters together, it is possible to preserve 
roughly the same stellar mass function as the reference model. This process also requires a 
slight adjustment to the AGN feedback threshold parameter, $\alpha_{\mathrm{cool}}$ to keep the 
break of the stellar mass function at the correct stellar mass. We refer to these
two variant models as the strong feedback (SFB, high mass loading, fast reincorporation) and
the weak feedback (WFB, low mass loading, slow reincorporation) models.

The other three models which we consider here are the three models presented in \cite{Mitchell14}.
For this paper, we are primarily interested in the two models from \cite{Mitchell14} that featured 
modified parametrisations for the reincorporation timescale. However, the models presented in 
\cite{Mitchell14} were run on merger trees extracted from the original Millennium simulation, 
which assumed a WMAP-1 cosmology \cite[][]{Springel05}. Therefore, to act as a point of comparison
for these two modified reincorporation models, we also include the reference model from \cite{Mitchell14}.
In this paper, we refer to the reference model from \cite{Mitchell14} as the M14 model.
In addition to the changes cosmological parameters, the three models taken 
from \cite{Mitchell14} also use the updated cooling scheme from \cite{Benson10b}. For reference,
the default \galform reincorporation timescale parametrisation (as used in the M14, reference, SFB 
and WFB models) is given by Eqn.~\ref{reincorporation} in Section~\ref{Simple_SFG_explanation_section}.

The first of the two modified reincorporation models from \cite{Mitchell14} which we consider
here, referred to as the star formation history (SFH) model, was designed
to try to reproduce the star formation histories for star forming galaxies inferred
from observations. For all but the most massive star forming galaxies, this model reproduces
the trend implied by observational data that the specific star formation rate at fixed stellar
mass has declined exponentially from high redshift to today. With respect to our reference
model, the SFH model uses a different parametrisation for the reincorporation rate, $\dot{M}_{\mathrm{ret}}$, given by

\begin{equation}
\dot{M}_{\mathrm{ret}} = \frac{\alpha_{\mathrm{reheat}}}{t_{\mathrm{dyn}}} \, \left(\frac{M_{\mathrm{H}}}{10^{10} h^{-1} \, \mathrm{M_\odot} } \right) \, F(z) M_{\mathrm{res}},
\label{reincorporation_modified_ch:msmh}
\end{equation}

\noindent where $t_{\mathrm{dyn}}$ is the halo dynamical time and $F(z)$ is a function of redshift given by

\begin{equation}
\log[F(z)] = 6 \, \exp{\left[-\frac{(1+z)}{3}\right]} \, \log_{10}[1+z].
\label{fudge_function_ch:msmh}
\end{equation}

\noindent This parametrisation has no physical motivation and essentially just represents an empirical
fit to the peaked star formation histories inferred for star forming galaxies in \cite{Mitchell14}.
This is achieved by making reincorporation rates very slow at early times when haloes have yet
to accrete most of their mass. The exponential function then dramatically lengthens the reincorporation
timescale at late times to achieve the exponential drop in star formation rates implied by 
observational data.

The final model from \cite{Mitchell14}, referred to here as the virial mass (VM) model, uses the 
reincorporation parametrisation advocated by \cite{Henriques13} and \cite{Henriques14}. This 
parametrisation is given by

\begin{equation}
\dot{M}_{\mathrm{ret}} = \alpha_{\mathrm{reheat}} \, \left(\frac{M_{\mathrm{H}}}{10^{10} h^{-1} \, \mathrm{M_\odot}}\right) \frac{M_{\mathrm{res}}}{1 \, \mathrm{Gyr}}.
\label{tret_bruno_ch:msmh}
\end{equation}

\noindent In appendix C of \cite{Mitchell14}, we showed that this model produces a 
good fit to the evolution in the stellar mass function below the break inferred from observations.

\begin{figure*}
\begin{center}
\includegraphics[width=40pc]{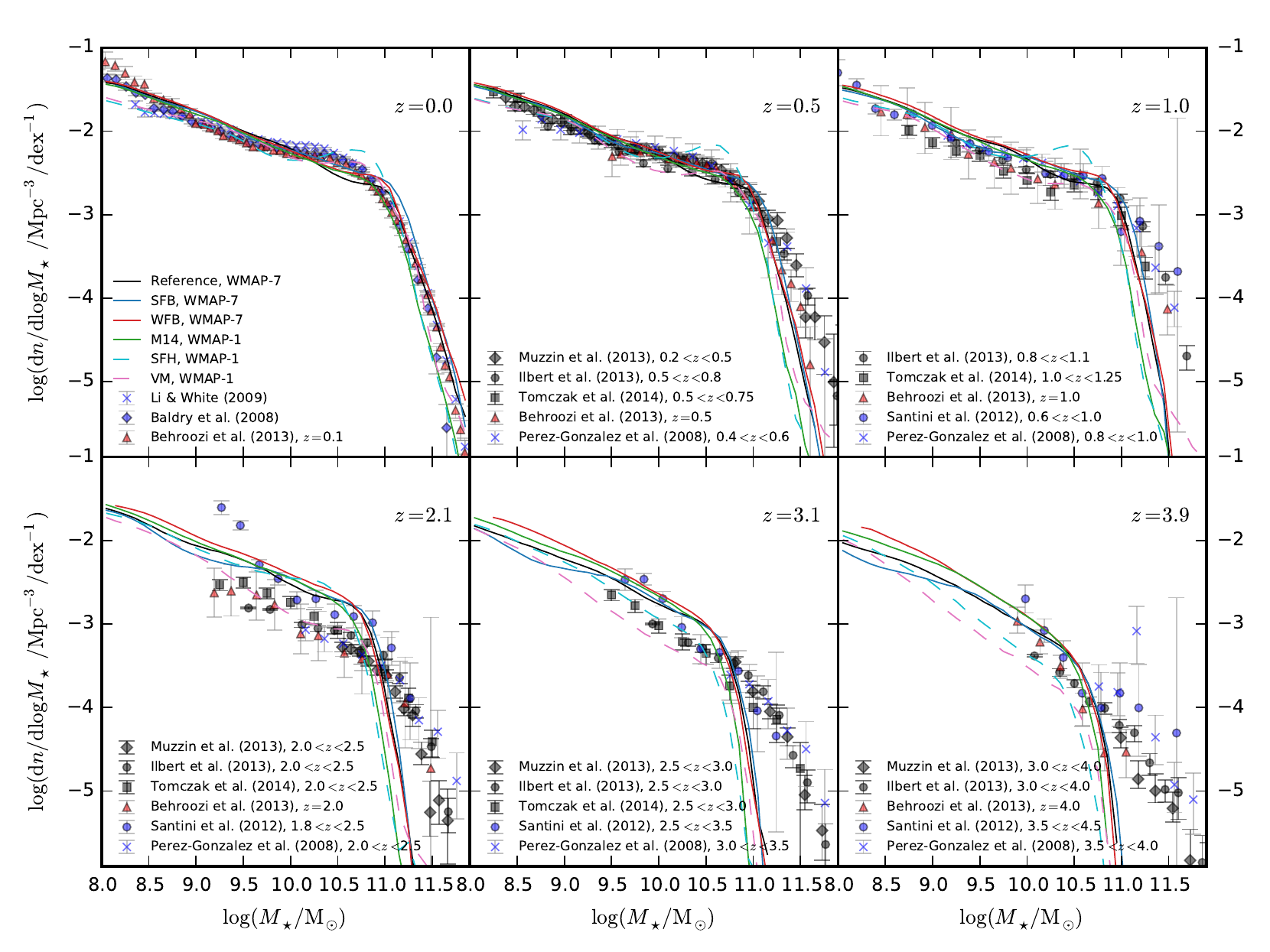}
\caption{Stellar mass functions for a selection of redshifts for the models described in \protect{Table~\ref{ModelsTable}}.
Each line corresponds to a different model, as labelled.
Points and associated errorbars show observational estimates of the stellar mass function
from \protect \cite{Baldry08}, \protect \cite{PerezGonzalez08}, \protect \cite{Li09}, \protect \cite{Santini12}, \protect \cite{Behroozi13}, \protect \cite{Ilbert13}, \protect \cite{Muzzin13} and \protect \cite{Tomczak14}.
Points that were used as constraints for abundance matching in \protect{Moster13} are shown in blue and those used by \protect \cite{Behroozi13} are shown in red.}
\label{smf_model_comp}
\end{center}
\end{figure*}

Before proceeding to analyse the predicted evolution in the SHM relation from the six models presented
in Table~\ref{ModelsTable}, we first show in Fig.~\ref{smf_model_comp} the stellar mass 
function for a range of redshifts for this family of models. At $z=0$, none of the models precisely match the shape of the
stellar mass function inferred from observations. Specifically, all models underpredict the abundance
of galaxies just below the knee. Furthermore, all but the VM and SFH models predict an overabundance of galaxies
at the low mass end (around $10^9$ to $10^{9.5} \, \mathrm{M_\odot}$). For this analysis however, we simply require that each model give a similar level
of agreement as the reference model to the observational estimates of the local stellar mass function.
As such, we consider the level of consistency between the models and data shown in the $z=0$ panel of 
Fig.~\ref{smf_model_comp} to be acceptable for our purposes. We note that in most instances the level
of disagreement between models and data is comparable to the level of disagreement between different
observational estimates.

\begin{figure*}
\begin{center}
\includegraphics[width=40pc]{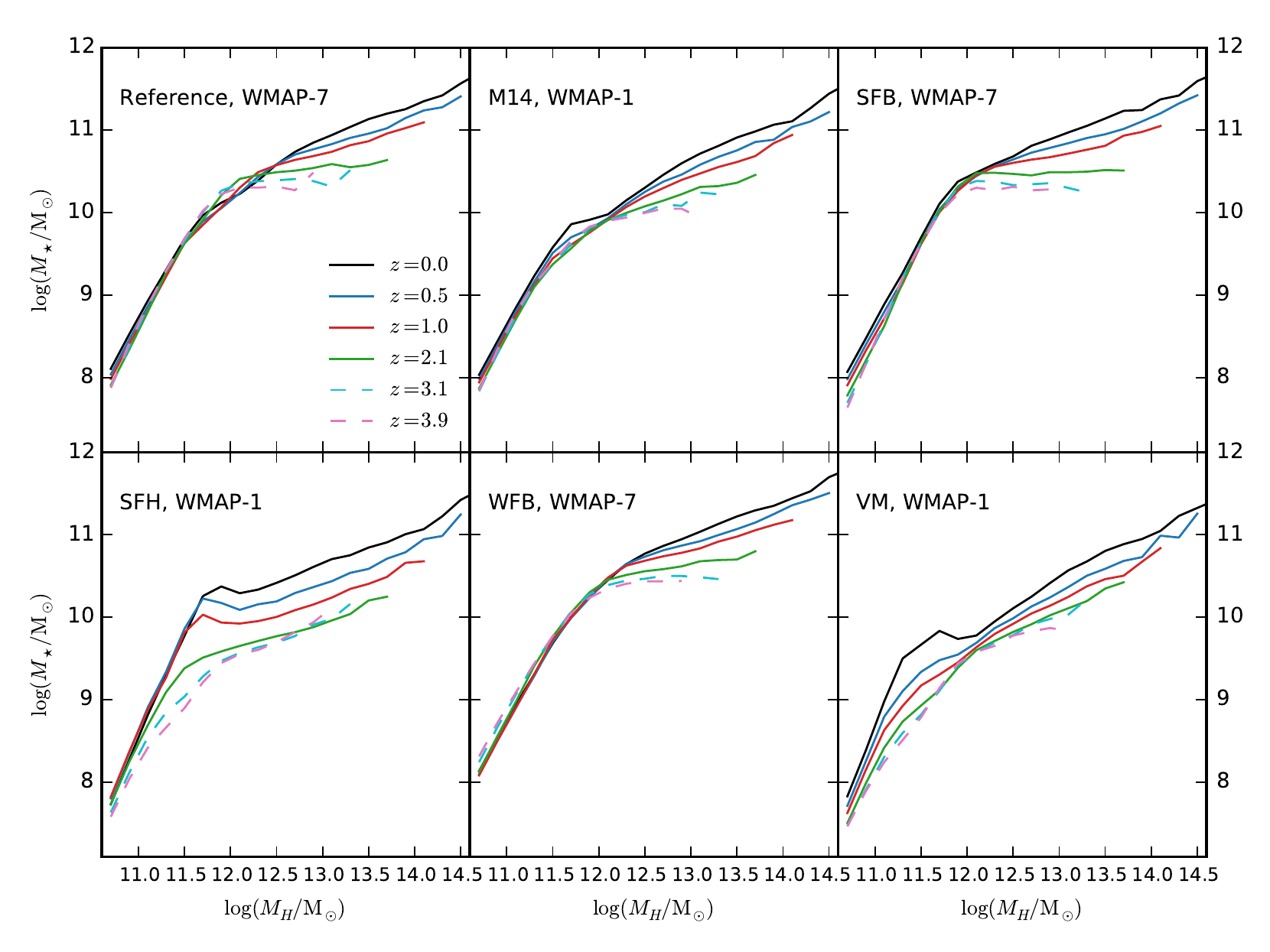}
\caption{Evolution of the median stellar mass as a function of halo mass for the models described in \protect{Table~\ref{ModelsTable}}.
Each panel corresponds to a different model, as labelled, while different lines within a panel show
the same model for different redshifts. The cosmological parameter set used for each
model is also labelled.
}
\label{smhm_model_comp}
\end{center}
\end{figure*}

Fig.~\ref{smhm_model_comp} shows the median SHM relation from the family of models presented in 
Table~\ref{ModelsTable} for a range of redshifts. Before proceeding to analyse the results, we
first note that when comparing the evolution predicted by different models, we expect the most 
prominent (and interesting) differences between the models will be displayed for halo masses 
both around and below the break in the SHM relation ($\log(M_{\mathrm{H}} / \mathrm{M_\odot}) < 12.5$). 
The reason for this expectation is that this is the halo mass range which contains star forming 
galaxies. In more massive haloes, stellar mass assembly takes place primarily through mergers, and 
the details of the SHM relation will be mainly determined by AGN feedback and the merging 
parametrisations, which we do not vary outside of adjusting the AGN feedback threshold, $\alpha_{\mathrm{cool}}$ 
\footnote{To first order, $\alpha_{\mathrm{cool}}$ can be considered as a parameter which only controls the break 
mass in the SHM relation.}. The variant models here are instead primarily distinct from each other 
in the parametrisations and parameters adopted for SNe feedback and gas reincorporation.
These are processes that mostly affect only the actively star forming galaxy population.

By examining Fig.~\ref{smhm_model_comp}, it is apparent that for halo masses above the SHM break, all the
models display similar (although not identical) evolution in the SHM relation. This presumably
reflects the fact that we do not change the AGN feedback model (beyond the threshold) or the
galaxy-galaxy merging timescale between the different models. In detail, the evolution should not be
(and is not) identical because, for example, of the role played by low mass satellites (which
are sensitive to the SNe feedback and gas reincorporation physics before infall) in building
the stellar mass of massive galaxies through minor mergers.

At and below the break 
($\log(M_{\mathrm{H}} / \mathrm{M_\odot}) < 12.5$), larger variations between some of the models become apparent.
Specifically, it can be seen that the trend for the SHM relation below the break to remain 
approximately constant with redshift is displayed by all the models (Reference, WFB, SFB, M14) 
using the standard reincorporation timescale. This is not an exact statement and the dynamic
range displayed in Fig.~\ref{smhm_model_comp} is large\footnote{The more subtle variations between the
Reference, WFB, SFB and M14 models are better viewed with lower dynamic range, which we address
with subsequent figures.}. Comparatively, the SFH and VM models display much more significant evolution
at and below the SHM break. For the SFH model in this halo mass range, the SHM relation evolves 
significantly for $z \ge 2$ before becoming fixed in place for $z \le 1$. This can be understood given that 
the model was designed implicitly to force star formation rates at fixed stellar mass to drop exponentially 
with cosmic time. The VM model also displays significant evolution in the SHM relation but in this case
the evolution also occurs for $z \le 1$. This behaviour can be understood because the VM model is designed 
implicitly to increase star formation rates at late times relative to the standard reincorporation
parametrisation used in the Reference, WFB, SFB and M14 models.

\begin{figure*}
\begin{center}
\includegraphics[width=40pc]{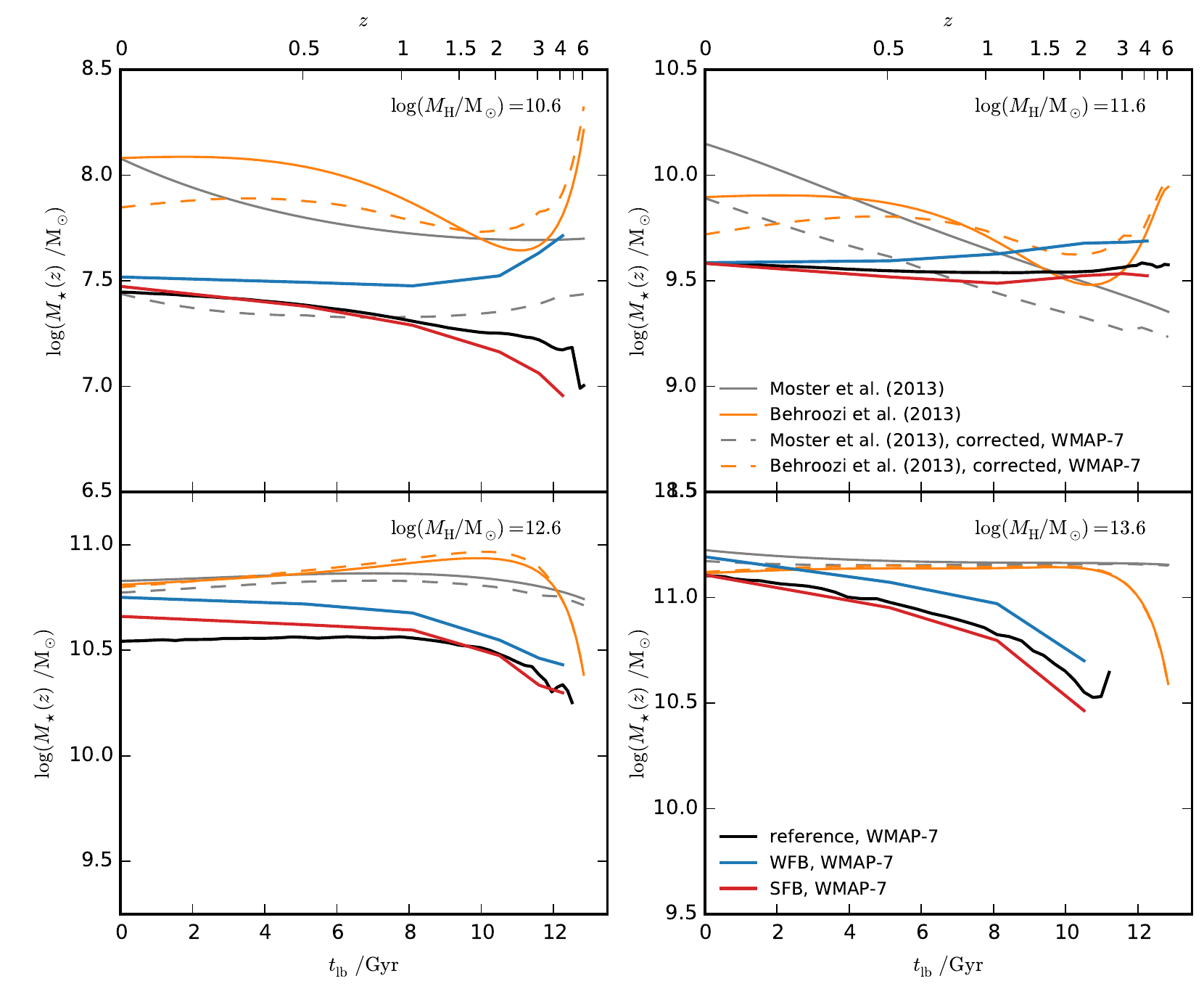}
\caption{Evolution in the median stellar mass within a given halo mass bin.
Each panel corresponds to a different halo mass bin, as labelled.
With the exception of grey and orange, solid lines show the median stellar mass for the subset of models described in \protect{Table~\ref{ModelsTable}} that use a WMAP-7 cosmology, as labelled.
Grey and brown solid lines show the best-fitting parametric SHM relations from \protect \cite{Moster13} and \protect \cite{Behroozi13} respectively.
\protect \cite{Moster13} and \protect \cite{Behroozi13} both assume a WMAP-7 cosmology.
Grey and brown dashed lines show best-fitting parametric evolution we obtain after correcting the \protect \cite{Moster13} and \protect \cite{Behroozi13} SHM relations to be compatible with the WMAP-7 halo catalogues used in \galform.
}
\label{mstar_mhalo_bin_model_comp2_wmap7}
\end{center}
\end{figure*}

\begin{figure*}
\begin{center}
\includegraphics[width=40pc]{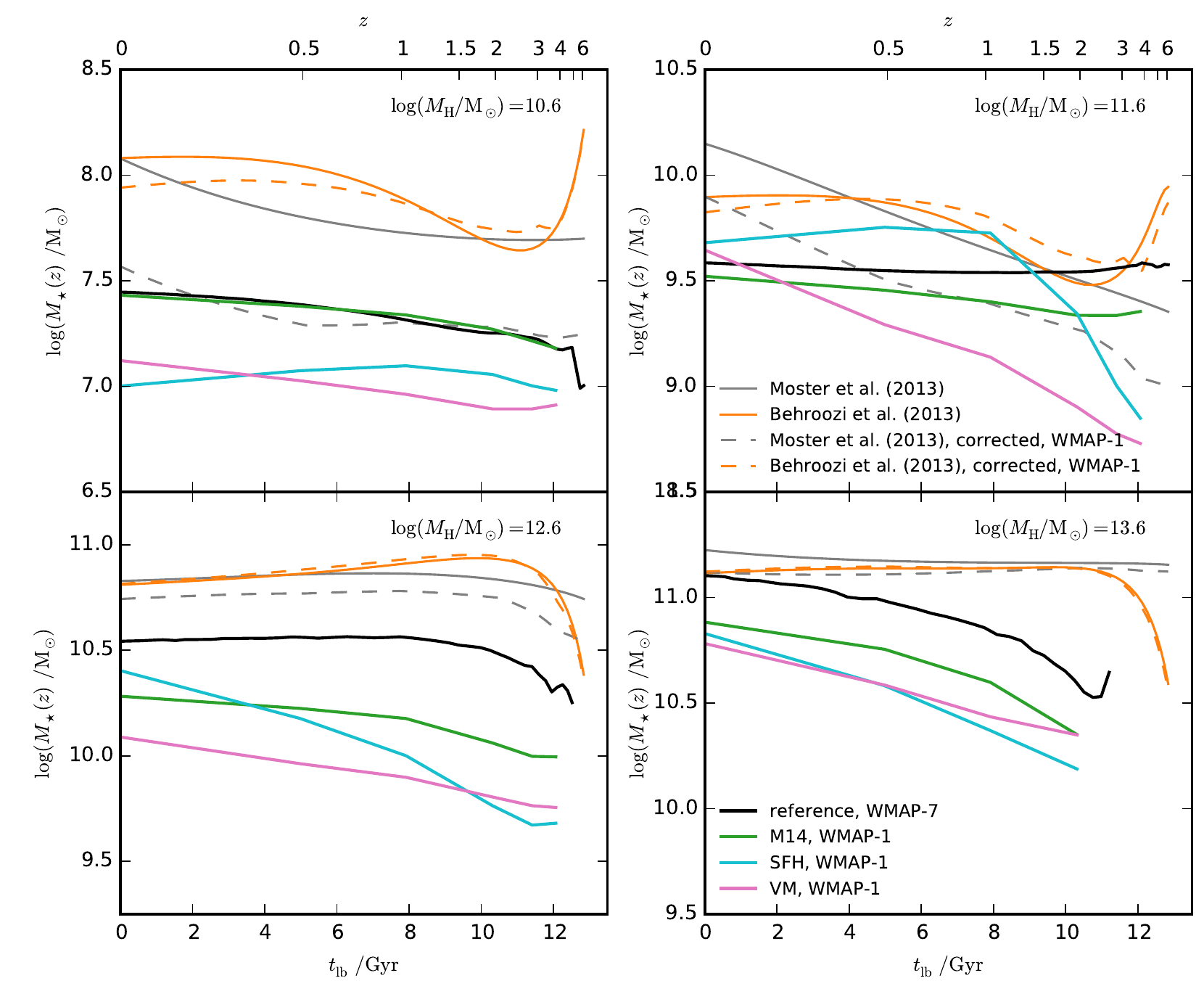}
\caption{Evolution in the median stellar mass within a given halo mass bin.
Each panel corresponds to a different halo mass bin, as labelled.
With the exception of grey and brown, solid lines show the median stellar mass for the subset of models described in \protect{Table~\ref{ModelsTable}} that use a WMAP-7 cosmology, as labelled.
Grey and orange solid lines show the best-fitting parametric SHM relations from \protect \cite{Moster13} and \protect \cite{Behroozi13} respectively.
\protect \cite{Moster13} and \protect \cite{Behroozi13} both assume a WMAP-7 cosmology.
Grey and brown dashed lines show best-fitting parametric evolution we obtain after correcting the \protect \cite{Moster13} and \protect \cite{Behroozi13} SHM relations to be compatible with the WMAP-1 halo catalogues used in \galform.
}
\label{mstar_mhalo_bin_model_comp2_wmap1}
\end{center}
\end{figure*}

Another view of the evolution of the SHM relation is shown in Fig.~\ref{mstar_mhalo_bin_model_comp2_wmap7} 
and Fig.~\ref{mstar_mhalo_bin_model_comp2_wmap1}, which show the evolution in median stellar mass at 
fixed halo mass. For these figures, we present the comparison with the SHM evolution inferred using 
abundance matching from \cite{Moster13} and \cite{Behroozi13}. We show the comparison both uncorrected 
(solid lines) and corrected (dashed lines) for differences with our halo catalogues (see 
Appendix~\ref{ap:msmh}). For clarity, as we are now considering a range of models with two different 
sets of cosmological parameters (WMAP-7 and WMAP-1), we have split the set of models from 
Table~\ref{ModelsTable} into two figures. The corrected abundance matching results shown in 
Fig.~\ref{mstar_mhalo_bin_model_comp2_wmap7} are therefore corrected to the \galform WMAP-7 halo 
catalogue while the corrected abundance matching results shown in Fig.~\ref{mstar_mhalo_bin_model_comp2_wmap1} 
are corrected to the \galform WMAP-1 halo catalogue.

Before commenting on the relative evolutionary trends displayed by the different \galform models shown
in Fig.~\ref{mstar_mhalo_bin_model_comp2_wmap7} and Fig.~\ref{mstar_mhalo_bin_model_comp2_wmap1}, 
it is worth underlining that the absolute discrepancies between
our models and abundance matching results in median stellar mass at a given halo mass are much
larger in some cases than might be expected from comparison of the stellar mass functions shown in 
Fig.~\ref{smf_model_comp}. In the lowest halo mass bin shown, the importance of the differences
in halo catalogues is clearly underlined as the \cite{Moster13} median stellar mass shifts down by
$\approx 0.4 \, \mathrm{dex}$, into better agreement with our \galform models. However, even
accounting for differences between halo catalogues, large differences remain. For example, the
median stellar mass at $\log(M_{\mathrm{H}} \, / \mathrm{M_\odot}) = 12.6$ in the VM model is 
$\approx 0.7 \, \mathrm{dex}$ lower than \cite{Moster13} and \cite{Behroozi13} at $z=0$. 
Given the fairly good agreement between the stellar mass functions in this model and the observational
constraints used by abundance matching at this redshift, and that the effects of halo catalogues
have been accounted for, this indicates that there are significant differences in the distribution of 
stellar mass around the median SHM relation.
Indeed, we find that the intrinsic scatter around the median SHM relation is significantly larger
in the SFH and VM models ($\sigma \approx 0.5-0.6 \, \mathrm{dex}$) compared to the other models with standard
gas reincorporation ($\sigma \approx 0.4 \, \mathrm{dex}$). It should also be noted that constraints inferred
from observations imply that the scatter should be significantly smaller, at $\sigma \approx 0.15-0.2 \, \mathrm{dex}$ (see
Appendix~\ref{ap:shm_distn}).

Returning our attention to the relative evolutionary trends shown by different models in 
Fig.~\ref{mstar_mhalo_bin_model_comp2_wmap7} and Fig.~\ref{mstar_mhalo_bin_model_comp2_wmap1}, 
we see that, as in Fig.~\ref{smhm_model_comp}, the 
VM and SFH models are clearly distinct from the other \galform models in that they predict 
significant evolution in the $\log(M_{\mathrm{H}}/\mathrm{M_\odot}) = 11.6$ bin. This is also 
the bin where the abundance matching results display the most significant evolution. Again, 
it is apparent that all of the models predict very similar evolutionary trends for the most 
massive haloes ($\log(M_{\mathrm{H}} /\mathrm{M_\odot}) = 13.6$ bin). It is interesting to note
that this trend seen in the models is contrary to abundance matching results which imply 
minimal evolution in this halo mass range. Given that the abundance matching results reproduce
  (by construction) the evolution of the stellar mass function inferred from observations, and
  that all of the models we consider here fail to reproduce the abundances of galaxies at the massive
  end of the stellar mass function at higher redshifts\footnote{This is not accounting for the Eddington bias effect,
    where due to steep shape of the Schecter function above the break, random stellar mass errors will preferentially up-scatter galaxies into the exponential tail of the distribution. This can lead to a significant overestimate of the abundance of
    galaxies above the exponential break in the stellar mass function.} 
  (see Fig.~\ref{smf_model_comp}), this implies that the models ought
  to be changed such that the SHM relation does not evolve at $\log(M_{\mathrm{H}} /\mathrm{M_\odot}) = 13.6$.

Fig.~\ref{mstar_mhalo_bin_model_comp2_wmap7} also shows more subtle differences between
the models. For example, in the $\log(M_{\mathrm{H}} /\mathrm{M_\odot}) = 10.6$ bin, the WFB and SFB models both 
clearly start to diverge from the reference model in opposite directions for $z \ge 1$. This
demonstrates how the degeneracy between $\alpha_{\mathrm{reheat}}$ and $V_{\mathrm{hot}}$
in the SHM relation  at $z=0$ in this halo mass range is broken by considering the evolution.

\begin{figure*}
\begin{center}
\includegraphics[width=40pc]{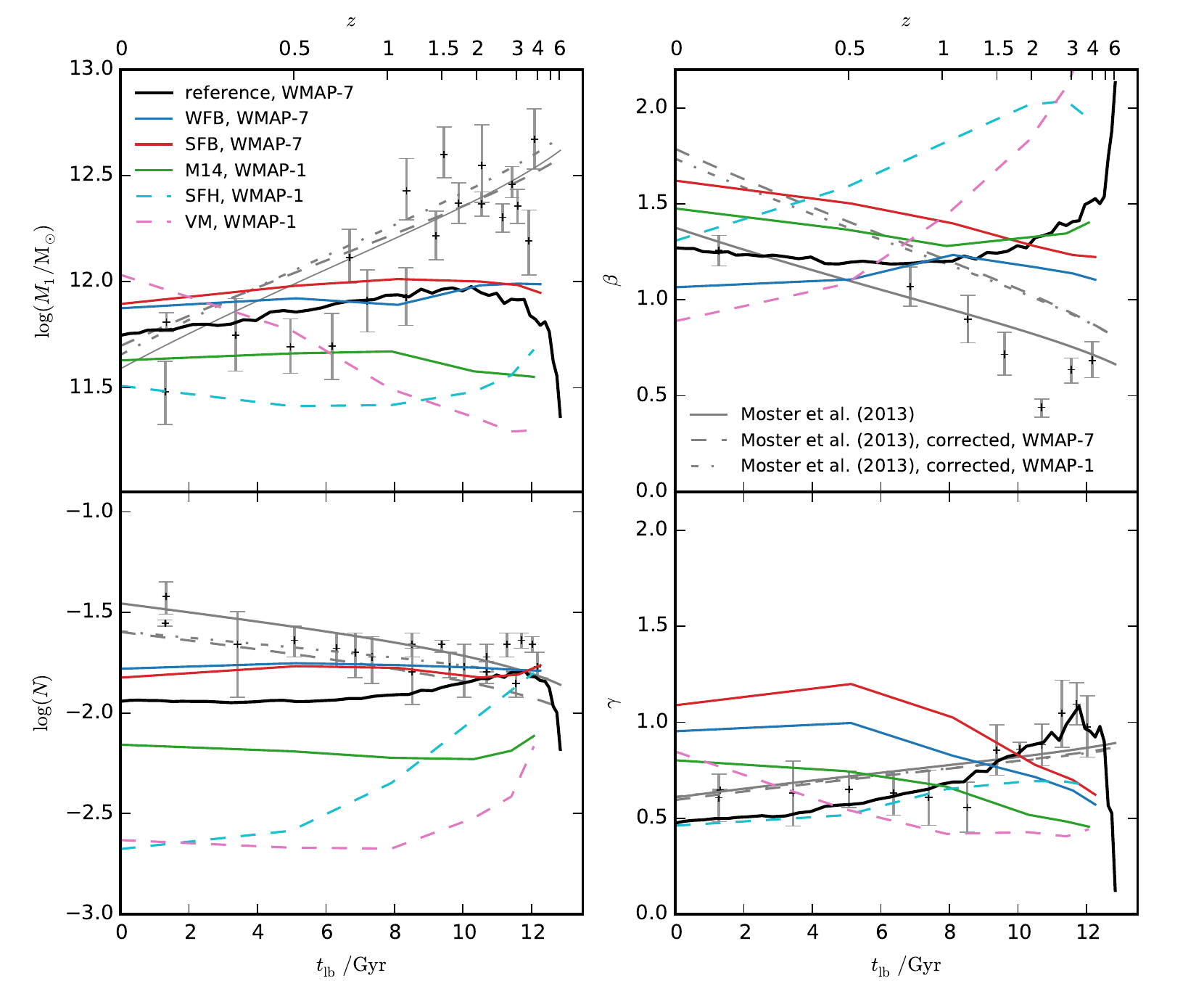}
\caption{Evolution in fitting parameters for the median SHM relation.
Lines (with the exception of grey) show the median of the projected posterior distribution for the models described in \protect{Table~\ref{ModelsTable}}, as labelled.
The cosmology used for each model is also labelled.
Grey solid lines show the best-fitting parametric evolution determined by \protect \cite{Moster13} using multi-epoch abundance matching, assuming a WMAP-7 cosmology.
Grey data points show the associated best-fitting SHM parameters and $ 1 \sigma$ errorbars determined by \protect \cite{Moster13} using single epoch abundance matching applied to individual stellar mass functions from the literature.
Grey dashed lines show the best-fitting parametric evolution we obtain after correcting the \protect \cite{Moster13} SHM relation to be compatible with the WMAP-7 halo catalogues used in \galform.
Grey dash-dotted lines show the same information but corrected to be compatible with WMAP-1 halo catalogues.}
\label{mstar_mhalo_fit_model_comp2}
\end{center}
\end{figure*}

An alternative view of the evolutionary behaviour in the SHM relation is presented in 
Fig.~\ref{mstar_mhalo_fit_model_comp2}, which shows the evolution in the fitting 
parameters from Eqn.~\ref{shm_eqn}. In this case, it is only possible to make the comparison
with the results from \cite{Moster13} (as we have adopted their parametrisation for the
SHM relation). We note that when considering results using the \cite{Moster13} SHM 
parametrisation given by Eqn.~\ref{shm_eqn}, it should be kept in mind that this parametrisation does not provide
a good fit to the SHM relations in the SFH and VM models at 
lower redshifts (see Fig.~\ref{smhm_model_comp}).

Starting with the break mass in the SHM relation, $M_{\mathrm{1}}$, 
Fig.~\ref{mstar_mhalo_fit_model_comp2} shows that the models we consider predict 
very little evolution. The exception is the VM model, which predicts that the break mass
drops by $\sim 0.7 \, \mathrm{dex}$ between $z=0$ and $4$. This is in contrast to
the trend inferred from \cite{Moster13}, who favour an increase in the break mass
towards high redshift. For the normalisation of the SHM relation at the break, $N$,
most of the models we consider predict minimal evolution, consistent with 
\cite{Moster13}. The exceptions are the VM and SFH models, where $N$ starts to
increase after $z=0.5$ and $z=1$ respectively.

For the low mass SHM slope, $\beta$, the differences between the different models become
more apparent. The WFB and SFB models again bracket the evolution predicted by the reference
model. The VM and SFH models predict that $\beta$ increases substantially with lookback
time, in contrast to the M14 model and \cite{Moster13}, demonstrating the importance of 
the reincorporation timescale parametrisation in galaxy formation models.
For the high mass SHM slope, $\gamma$, the models we consider all predict fairly
modest evolution, consistent with \cite{Moster13}.

\section{Discussion}
\label{discussion}

As a diagnostic of galaxy formation models, comparing the median 
SHM relation predicted by competing models with the results of abundance matching 
can provide complementary information to a comparison between model 
predictions and observational estimates of the stellar mass function.
If a given galaxy formation model reproduces the observational estimates of the stellar
mass function that were used to constrain an abundance matching model,
then any differences in the predicted versus inferred median SHM relation
can be interpreted as a problem in the galaxy formation model with the distribution in 
stellar mass around the median SHM relation. The two caveats to this
are firstly that the halo catalogues used as inputs for the two techniques
must be the equivalent. Secondly, the abundance matching itself needs to 
adequately reproduce the true intrinsic scatter around the median SHM 
relation (and model potential sources of error on observational data correctly).
In Section~\ref{parameters_section}, we highlighted several instances where the median SHM relation
inferred using abundance matching was discrepant with specific \galform models,
despite correcting for differences in halo catalogues. Given that some of these
models give reasonable levels of agreement with the observational estimates of
the stellar mass function used to constrain abundance matching, one interpretation
therefore has to be that the distribution in stellar mass around the 
median in our models is not the same as for real galaxies.

In \cite{Mitchell14} we found that it was necessary to modify the parametrisation
of at least one of the physical processes in our model in order to reproduce the
star formation rates of star forming galaxies inferred from observations.
As one of the most uncertain aspects of our modelling approach, we chose to modify the 
reincorporation timescale to illustrate this point. However, we then
found that explaining the evolution of the stellar mass function requires a contradictory
modification to the gas reincorporation timescale compared to explaining the evolution
of star formation rates. 
Specifically, we introduced the SFH model to reproduce the star formation rate evolution
inferred from observations and the VM model to reproduce the evolution of the stellar mass function\footnote{These models are introduced in Section~\ref{AlternativeSection}.}.
Given that these modifications to the reincorporation timescale have a significant impact
on the predicted stellar mass functions and star formation rates, one naturally expects differences
to also appear in the predicted evolution of the SHM relation. We find that this is indeed
the case close to the break in the SHM relation. However, neither of our modified models 
predict evolution that closely resembles results from the abundance matching studies of \cite{Moster13} 
and \cite{Behroozi13}, despite claims from those studies that they reproduce simultaneously 
both the star formation rates and the stellar mass assembly inferred from observations.
This is interesting, particularly given the problems that have been reported by a wide range
of contemporary models and simulations in reproducing evolution in star formation rates and/or 
stellar mass functions inferred from observations 
\cite[e.g.][]{Lamastra13,Furlong14,Cousin15,Sparre15}, but see \cite{Henriques14}.

As we have discussed, this could reflect problems with the intrinsic distribution
of stellar mass around the median stellar mass at a given halo mass. We plan to
return to this topic as part of future work (see also the discussion in 
Appendix~\ref{ap:shm_distn}). However, there are other aspects of the models and 
abundance matching that are worthy of consideration. When considering the problem
of reproducing star formation rates and stellar mass functions simultaneously,
it is important to be aware that the abundance matching approach does not, 
at present, distinguish between star forming and passive galaxy populations at a 
given halo mass \cite[although see][]{Hearin13,Watson14}. This is likely to be 
problematic close to the SHM break mass, where the central galaxy population 
transitions between dominant star forming and passive galaxy populations. We note 
that this is precisely the most interesting halo mass range for distinguishing between 
the different models considered in our analysis.

Another important consideration is whether recent observational estimates of the stellar mass 
function from deep Ultra-VISTA \cite[][]{Ilbert13,Muzzin13} and ZFOURGE \cite[][]{Tomczak14} 
data display significant differences with respect to older observational estimates, particularly 
above $z=2$. From Fig.~\ref{smf_model_comp}, where these
recent estimates can be compared with the estimates used as abundance matching constraints,
we conclude that the constraints used by \cite{Moster13} and \cite{Behroozi13} are not
obviously in significant disagreement with the more recent observational estimates.
At higher redshifts, the \cite{Santini12} estimate \cite[used as a constraint in][]{Moster13}
is perhaps too steep at lower stellar masses, and the estimate from \cite{Behroozi13}
for the stellar mass function at $z=2$ is perhaps a little low around the knee. However,
we would not expect the estimates of the median SHM relation from abundance matching 
(or other techniques) to change significantly once these more recent datasets are 
included as constraints.

\section{Summary}
\label{summary}

We have explored the evolution of the median stellar mass versus halo mass (SHM) relation predicted by different versions
of the semi-analytic galaxy formation model, \galform. For our reference model, where the return
timescale for gas ejected from galaxies by SNe feedback scales with the halo dynamical
timescale, we find that the median SHM evolves only very modestly between $z=0$ and $z=4$.
This implies that the efficiency of stellar mass assembly (star formation plus galaxy mergers)
within haloes at fixed halo mass is approximately independent of cosmic time \cite[see][for a discussion
of this point]{Behroozi13b}. 
In our model, this behaviour is primarily driven by the evolution of the efficiency of SNe feedback
in regulating star formation rates of actively star forming galaxies. This SNe efficiency drops
as haloes grow in mass, such that star forming galaxies evolve along a fixed power law 
in the stellar mass versus halo mass plane, given by $M_\star \propto M_{\mathrm{H}}^{2.3}$. Another 
factor that causes there to be minimal evolution in the predicted SHM relation is the AGN feedback 
model implemented within \galform. Specifically, we show that the threshold for 
AGN feedback to become effective at suppressing gas cooling in haloes corresponds to 
a halo mass which is only weakly dependent on cosmic time. This causes the break mass in the SHM 
relation predicted by \galform to evolve only very modestly.

To reproduce the shape of the local stellar mass function inferred from observations places
strong constraints on the form of the median SHM relation at $z=0$. We show that with this single constraint
in place, standard\footnote{By this we mean models where the efficiency of gas reincorporation 
after ejection by feedback evolves only weakly with cosmic time.} semi-analytic galaxy formation 
models tend not to predict significant evolution in the SHM relation. 
This behaviour is broken close to the SHM break mass (which is closely
connected to the knee of the stellar mass function) for the models introduced in \cite{Mitchell14}
that feature modified gas reincorporation timescales. At present, abundance matching studies
\cite[][]{Behroozi13,Moster13} do not strongly support either of these modified \galform
models. Our preliminary interpretation of this disagreement is that there could likely be a problem
in our modified models with, at a given halo mass, the form of the distribution in stellar mass 
around the median SHM relation. We conclude therefore that there is a clear opportunity to use constraints
from the full distribution in stellar mass as a function of halo mass, inferred using empirical 
modelling of observational data, to improve theoretical galaxy formation models.

\section*{Acknowledgements}

We thank Qi Guo for making catalogues from the L-galaxies semi-analytic galaxy formation model available to us.
We thank Idit Zehavi for reading and providing comments that helped improve the clarity of this paper.
This work was supported by the Science and Technology Facilities Council [ST/J501013/1, ST/L00075X/1].
This work used the DiRAC Data Centric system at Durham University, operated by the Institute for Computational Cosmology on behalf of the STFC DiRAC HPC Facility (www.dirac.ac.uk). 
This equipment was funded by BIS National E-infrastructure capital grant ST/K00042X/1, STFC capital grant ST/H008519/1, and STFC DiRAC Operations grant ST/K003267/1 and Durham University. 
DiRAC is part of the National E-Infrastructure.

\bibliographystyle{mn2e}
\bibliography{bibliography}

\appendix

\input{appendix_msmh}

\input{appendix_msmh3}

\input{appendix_msmh2}

\input{appendix_msmh4}

\label{lastpage}
\end{document}

%% file: appendix_msmh.tex
\section{Halo masses and satellite abundances}
\label{ap:msmh}

Here, we describe how we attempt to account for differences in the definition of halo mass and in the abundance of satellites
between the \galform model and the abundance matching studies of \cite{Moster13} and \cite{Behroozi13}. In brief, we match the input halo
catalogues used by these two studies to correct their reported SHM relations to be consistent with the halo catalogues used in \galform.

To do this, we require realisations of the halo catalogues used as inputs by \cite{Moster13} and \cite{Behroozi13}. For the case of \cite{Moster13},
we also need to know the distribution of infall redshifts for satellite galaxies. For this purpose, we have made use of an L-galaxies model which was run 
using the same {\sc{MR7}}\xspace simulation that was used in our reference model (Guo, private communication).
Compared to \galform, the L-galaxies model uses the same SUBFIND subhalo catalogues as inputs but uses halo mass definitions 
and assumptions about satellites which are the same as those adopted by \cite{Moster13}. Specifically, L-galaxies uses a mean halo 
density of $200$ times the critical density ($M_{\mathrm{200}}$) to define halo mass, and a dynamical friction timescale is used to 
decide how long satellite galaxies survive after their subhalo can no longer be identified in the simulation \cite[][]{Guo11,Moster13}.

For the case of \cite{Behroozi13}, we simply generate a Monte-Carlo realisation of their halo catalogue using the corrected \cite{Tinker08} 
halo mass function and the satellite fractions reported in appendix C of \cite{Behroozi13}. \cite{Behroozi13} define halo masses using the 
virial overdensity criterion predicted by the spherical collapse model \cite[][]{Bryan98}. We note that unlike \cite{Moster13}, 
\cite{Behroozi13} do not include a population of orphan satellites (satellite galaxies with no identifiable subhalo in the simulation),
which we will refer to as Type 2 satellites. We refer to satellites for which the subhalo can still be identified in the simulation as Type 1 
satellites.

\begin{figure}
\includegraphics[width=20pc]{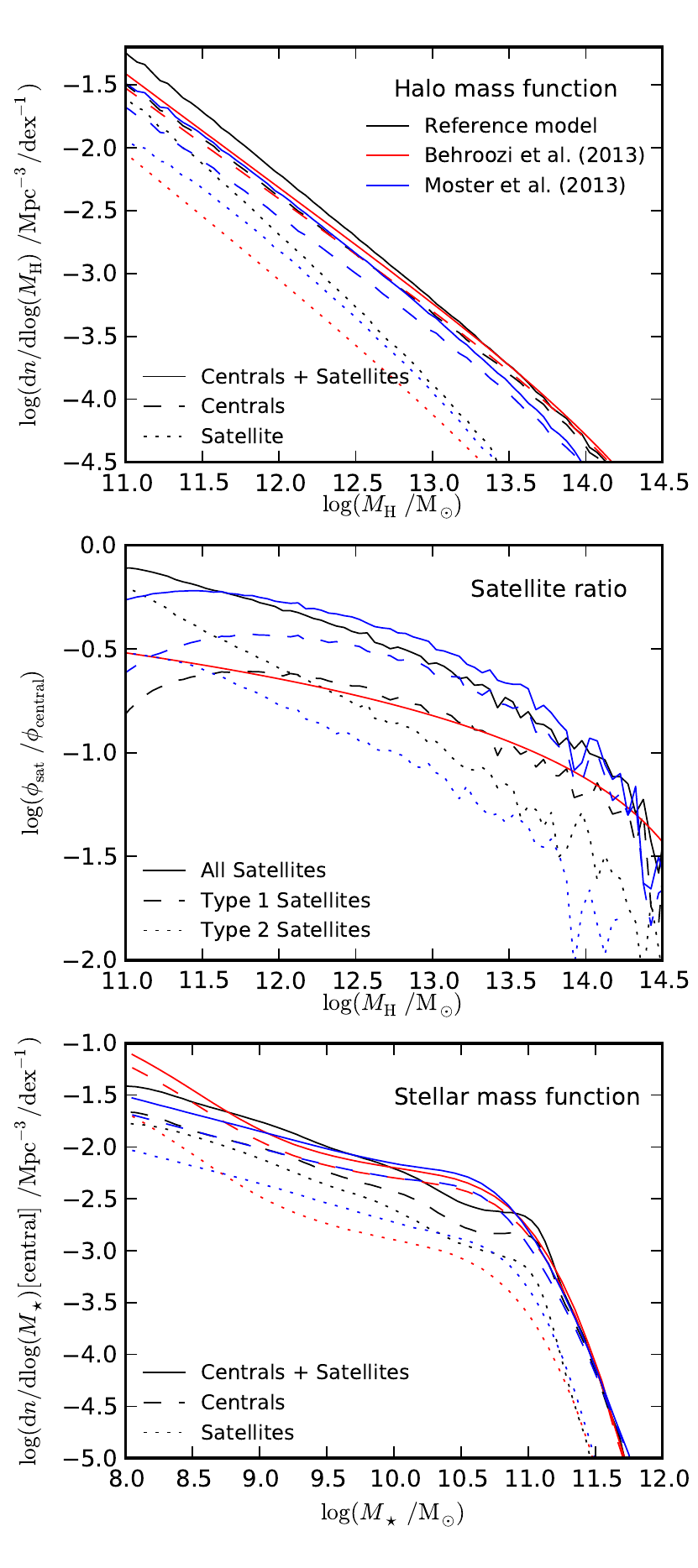}
\caption{Mass functions at $z=0$ from our reference \galform model (black) and from \protect \cite{Moster13} (blue) and \protect \cite{Behroozi13} (red).
{\it Top:} Halo mass functions, using the halo mass definitions used by each study to quantify the SHM relationship.
Solid lines show combined central plus satellite halo mass functions, using the subhalo mass at infall for satellites.
Dashed lines and dotted lines show the halo mass functions for central and satellite galaxies respectively.
{\it Middle:} Ratio of the satellite halo to the central halo mass functions, as a function of halo mass.
Solid lines show the ratio for all satellites. Dashed lines and dotted lines show the ratio for Type 1 and Type 2 satellites respectively.
{\it Bottom:} Stellar mass functions for all galaxies (solid lines), central galaxies (dashed lines) and satellite galaxies (dotted lines, includes both Type 1 and Type 2 satellites).
}
\label{abundance_matching_comp}
\end{figure}

The importance of accounting for the different halo catalogues used by our model, \cite{Moster13} and \cite{Behroozi13} is illustrated
in Fig.~\ref{abundance_matching_comp}. We show the halo mass functions, satellite-to-central ratios and the resulting stellar mass 
functions from the three models at $z=0$. The central halo mass functions from \galform and \cite{Behroozi13} are almost indistinguishable
(they use similar halo mass definitions) while the \cite{Moster13} central halo mass function is systematically offset to lower halo mass at a given number
density. The offset in $\log(M_{\mathrm{H}} / \mathrm{M_\odot})$ is $\approx 0.16 \, \mathrm{dex}$. 

For satellite galaxies, there is a reasonable agreement in the ratio of satellites (Type 1 plus Type 2) to centrals between \galform and \cite{Moster13}.
There is however a difference in the relative fractions of Type 1 compared to Type 2 satellites between the two halo catalogues. This primarily 
reflects the fact that in \galform, all satellites are allocated an analytically calculated dynamical friction merging timescale at infall, instead of when the
subhalo is lost from the simulation, as is the case in \cite{Moster13}. For the \cite{Behroozi13} halo catalogue, the satellite to central ratio
is significantly smaller than in the \galform or \cite{Moster13} halo catalogues. This presumably reflects in part the decision made by \cite{Behroozi13} not to include
Type 2 satellites.

The net result of the difference in satellite abundances is reflected in the combined central plus satellite halo mass function (solid
lines in the top panel of Fig.~\ref{abundance_matching_comp}). We emphasise that it is the combined halo mass function that is relevant
for abundance matching. The \cite{Moster13} combined halo mass function is similar to the \galform halo mass function, but with a roughly
constant offset in halo mass at fixed number density. The \cite{Behroozi13} combined halo mass function
agrees with the \galform combined mass function for massive haloes (where centrals dominate) but is shallower. 
This is caused by the smaller abundance of satellites
in the \cite{Behroozi13} catalogue. We also show the resulting stellar mass functions in the bottom panel of Fig.~\ref{abundance_matching_comp}.
For \cite{Moster13} and \cite{Behroozi13}, these are produced using their halo catalogues populated using their SHM relations, including intrinsic scatter.

To account for the differences between halo catalogues, we attempt to find a way to correct the \cite{Moster13} and \cite{Behroozi13} SHM relations
such that they resemble the SHM relations that they would have obtained if they had performed abundance matching using the \galform halo catalogue.
Specifically, we search for appropriate mapping functions, $F(M_{\mathrm{H}},z)$, that correct halo masses from the \cite{Moster13}
and \cite{Behroozi13} halo catalogues (such that $M_{\mathrm{H,corrected}} = F(M_{\mathrm{H}}) M_{\mathrm{H}}$). Corrected SHM relations are then obtained
by applying their SHM relations to the corrected halo catalogues. The redshift dependence of $F(M_{\mathrm{H}},z)$ reflects that different mapping functions,
$F(M_{\mathrm{H}})$, will be required to correct the SHM relation for different redshifts. In the case of \cite{Moster13}, we note that for satellites
we use a correction factor, $F(M_{\mathrm{H}},z_{\mathrm{infall}})$, that corresponds to the infall redshift of the satellite, $z_{\mathrm{infall}}$.

To find $F(M_{\mathrm{H}},z)$, we proceed as follows. As a starting point, if we were to consider, for example, the halo catalogue from 
\cite{Behroozi13} at a given redshift, we can apply their SHM relation to both the \galform halo catalogue and their catalogue to obtain two cumulative stellar mass functions.
If the two halo catalogues differ, then the resulting stellar mass functions will also differ. Our task is to find $F(M_{\mathrm{H}},z)$ such that by applying 
$F(M_{\mathrm{H}},z)$ as a correction to the halo masses in the Behroozi et al. catalogue before applying their SHM relation to compute a stellar
mass function, the resulting cumulative stellar mass functions from the \galform and Behroozi et al. halo catalogues are 
equal\footnote{Note therefore that this target stellar mass function is not the stellar mass function predicted by our \galform model or the stellar mass function from the empirical \cite{Behroozi13} model.}.
We use this as our constraint because \cite{Moster13} and \cite{Behroozi13} match abundances as a function of stellar mass\footnote{We neglect the fact that \cite{Behroozi13}
also use other observational constraints to constrain their model parameters.}.

At a given redshift, we find that a double power law for $F(M_{\mathrm{H}})$ is appropriate, with a form given by

\begin{equation}
F(M_{\mathrm{H}}) = N_{\mathrm{r}}  \, \left[ \left(\frac{M_{\mathrm{H}}}{M_{\mathrm{r}}}\right)^{\alpha_{\mathrm{r}}} + \left(\frac{M_{\mathrm{H}}}{M_{\mathrm{r}}} \right)^{\beta_{\mathrm{r}}} \right],
\label{remap_eqn}
\end{equation}

\noindent where $N_{\mathrm{r}}$, $M_{\mathrm{r}}$, $\alpha_{\mathrm{r}}$ and $\beta_{\mathrm{r}}$ are fitting parameters. Once
we obtain $F(M_{\mathrm{H}})$ for a set of redshifts, we estimate $F(M_{\mathrm{H}},z)$ simply by interpolating 
$\log[F(M_{\mathrm{H}})]$ in $\log(1+z)$. We note that we choose not to extrapolate $F(M_{\mathrm{H}})$ for halo
masses larger than we can constrain using our halo catalogues at a given redshift. Instead we hold constant $F(M_{\mathrm{H}})$ above this mass.

To estimate values for the fitting parameters in Eqn.~\ref{remap_eqn}, we can obtain a first guess simply by directly matching
abundances between two halo catalogues (one from \galform and one from either of Moster et al. or Behroozi et al.) as a function of halo mass
(instead of stellar mass). For the case of \cite{Behroozi13}, this first step is all that it is required because in their empirical model,
the stellar masses of galaxies depend only on their SHM relation evaluated at the redshift of interest. For the case of \cite{Moster13}, 
there is an additional complication because they assign satellite galaxies with a
stellar mass drawn from the SHM distribution that corresponds to the infall redshift of each satellite, rather than from the 
distribution corresponding to the desired redshift. Therefore, the abundance of galaxies as a function of 
stellar mass depends on more than just the halo catalogue at the output redshift. In this case, we have to employ a minimisation procedure
to refine our intial guess for the fitting parameters in Eqn.~\ref{remap_eqn}.

\begin{figure}
\includegraphics[width=20pc]{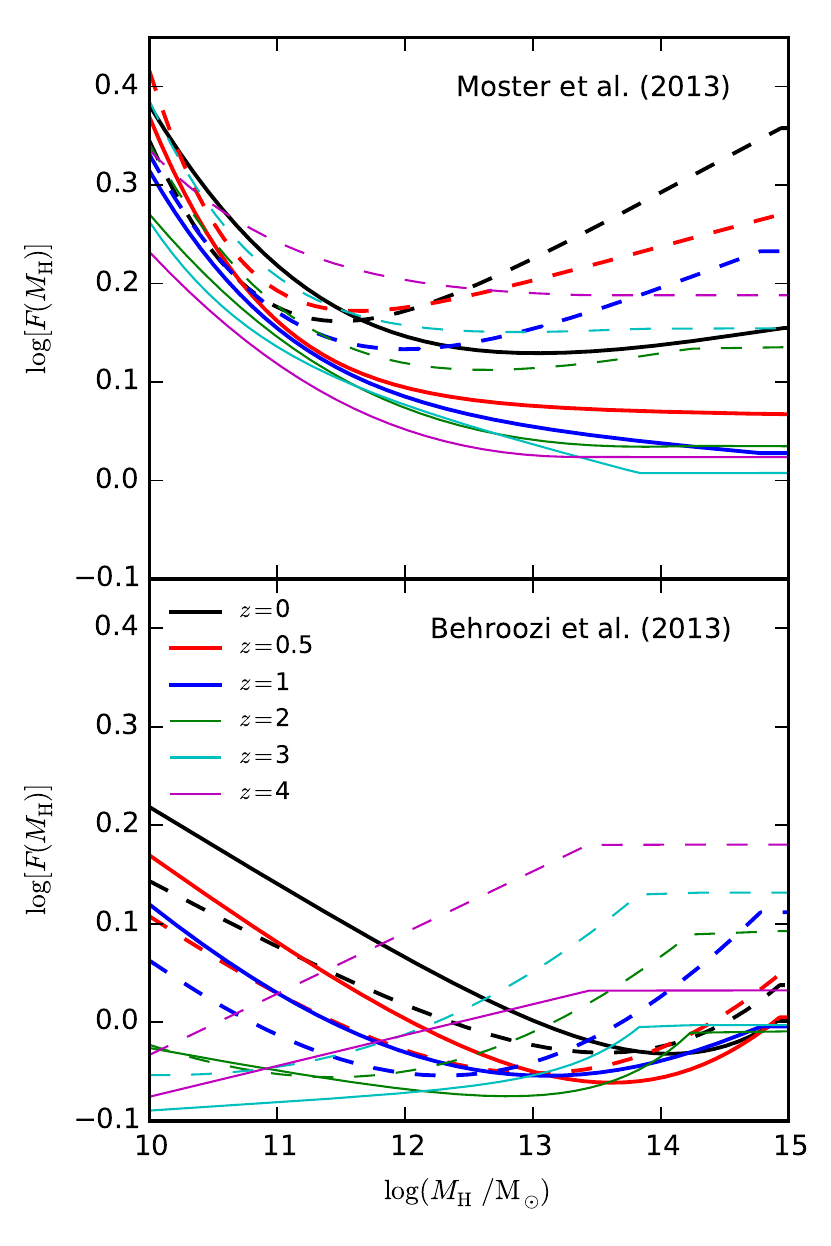}
\caption{Function $F(M_{\mathrm{H}})$ that maps between the halo masses from two different halo catalogues such that, for a given SHM relation, the two catalogues give the same total abundance of galaxies as a function of stellar mass.
The top panel shows $F(M_{\mathrm{H}})$ computed between the halo catalogues used in \galform and a catalogue that is effectively identical to the one used by \protect \cite{Moster13}.
Solid lines show $F(M_{\mathrm{H}})$ for the \galform catalogue taken from the {\sc{MR7}}\xspace simulation.
Dashed lines show $F(M_{\mathrm{H}})$ for the \galform catalogue taken from the {\sc{millennium}}\xspace simulation.
Different coloured lines correspond to different redshifts, as labelled.
The bottom panel shows the corresponding $F(M_{\mathrm{H}})$ computed between the halo catalogues used in \galform and a Monte-Carlo realisation of the catalogue used by \protect \cite{Behroozi13}.
}
\label{remap_halo_mass}
\end{figure}

For the results presented in the main body of this paper, we have computed the mapping functions 
appropriate for converting SHM relations to be compatible with either our {\sc{MR7}}\xspace (WMAP-7 cosmology) or {\sc{millennium}}\xspace
(WMAP-1 cosmology) simulations. We show $F(M_{\mathrm{H}}, z)$ as a function of halo mass for the different simulations and
abundance matching studies in Fig.~\ref{remap_halo_mass}.

%% file: appendix_msmh3.tex
\section{Intrinsic scatter in the local SHM distribution}
\label{ap:shm_distn}

\begin{figure*}
\begin{center}
\includegraphics[width=40pc]{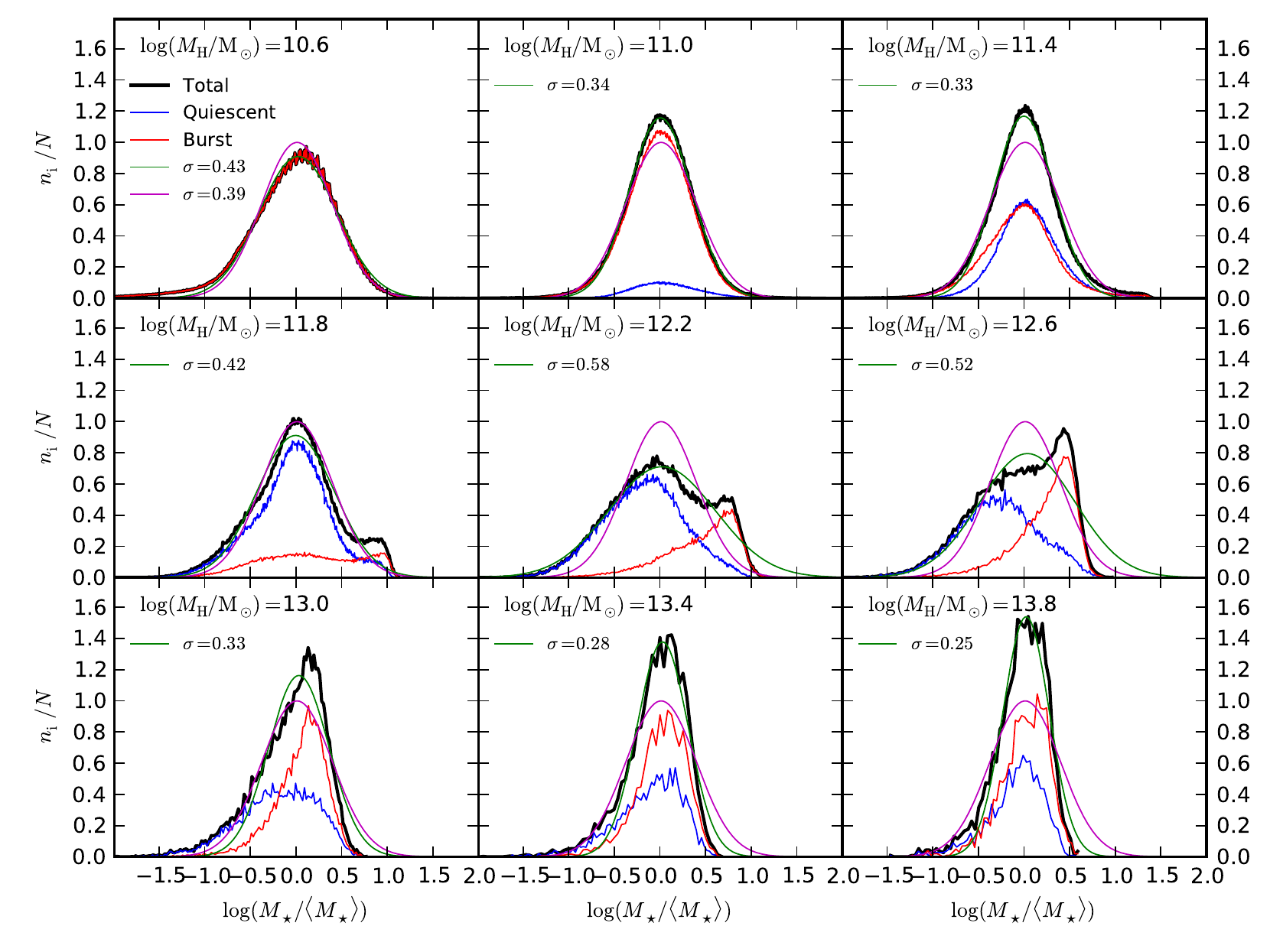}
\caption{Distributions of stellar mass for the total galaxy population, split into narrow ($\Delta \log(M_{\mathrm{H}}/\mathrm{M_\odot}) = 0.1 \, \mathrm{dex}$) bins in halo mass from our reference model at $z=0$ (black lines).
Each panel corresponds to a different halo mass bin, as labelled.
For each halo mass bin, we divide stellar masses through by the median stellar mass of the bin.
We then fit lognormal distributions individually to each bin (green lines) and to all bins simultaneously (magenta lines).
We also show distributions separated on the basis of whether the stellar mass within a given galaxy spheroid was assembled primarily through bursts of star formation (red lines) or through 
quiescent star formation in galaxy discs which was subsequently added to the spheroid through mergers or disc instabilities (blue lines).
}
\label{shm_hist}
\end{center}
\end{figure*}

Here, we present the distributions in stellar mass for a set of narrow ($\Delta \log(M_{\mathrm{H}} / \mathrm{M_\odot}) = 0.1 \, \mathrm{dex}$) bins in halo mass
from our reference model. These are shown in Fig.~\ref{shm_hist}. Our aim is to illustrate that our reference
model does not predict that the SHM distribution at fixed halo mass is strictly lognormal, with constant width as a function of halo mass, as is often assumed.
To demonstrate this we fit lognormal distributions both individually to each bin (green lines) and to all bins simultaneously (magenta lines).
For low ($M_{\mathrm{H}} < 10^{11.5} \, \mathrm{M_\odot}$) and high ($M_{\mathrm{H}} > 10^{13.0} \, \mathrm{M_\odot}$) mass haloes, a lognormal 
distribution appears to provide a good description of the SHM distribution. However, in the intermediate halo mass range, the SHM distribution 
at fixed halo mass is skewed with respect to a lognormal. Furthermore, from the lognormal distributions that are fit individually to each bin, 
it is apparent that the intrinsic scatter in the SHM distribution is a function of halo mass, with the scatter peaking at a halo mass of 
$M_{\mathrm{H}} \approx 10^{12.2} \, \mathrm{M_\odot}$.

In the $10^{12.2} \, \mathrm{M_\odot}$ halo mass bin, the distribution is visibly bimodal. We find that
this bimodality is best explained by splitting the galaxy population according to how galaxy spheroids assembled their stellar mass. We first consider galaxies
where the majority of the stellar mass in the galaxy spheroid was formed as part of quiescent star formation in discs which were subsequently
added to the spheroid through either galaxy mergers or disc instabilities. The second population we consider is instead comprised of galaxies where the majority
of the stellar mass in the galaxy spheroid was formed in bursts of star formation that took place within galaxy spheroids. In our model, these bursts
are triggered by gas accretion onto a spheroid during galaxy merger or disc instability events.
Fig.~\ref{shm_hist} shows that in intermediate mass haloes ($\approx 10^{12.6} \, \mathrm{M_\odot}$), the galaxies with spheroids where the
stellar mass was originally assembled by quiescent disc star formation (blue lines) have total stellar masses that are systematically lower
than galaxies where the spheroids were assembled in bursts.

To explain this bimodality we remind the reader that in \galform, the mass loading factor for SNe feedback is parametrised as a function of 
disc circular velocity for star formation in galaxy discs, and as a function of spheroid circular velocity for star formation taking
place in nuclear bursts.
  In the halo mass range where SNe feedback is strong ($M_{\mathrm{H}} < 10^{13} \, \mathrm{M_\odot}$), any systematic differences
  between typical disc and spheroid circular velocities leads to significantly different mass loading factors for SNe feedback  between quiescent and burst star formation channels (which are amplified because of the exponent in the mass loading parametriation, $\alpha_{\mathrm{hot}} = 3.2$). This means that at a given
halo mass, the efficiency of star formation will depend sensitively on whether star formation takes place in bursts or quiescently in discs. Therefore, in the halo mass
range where SNe feedback is strong and the spheroid mass can be significant fraction of the total stellar mass of a given galaxy
($10^{11.8} < M_{\mathrm{H}} \, \mathrm{M_\odot} < 10^{13}$), there can be significant differences between $M_{\star} / M_{\mathrm{H}}$ at a given halo mass
depending on whether bursts or quiescent star formation were the dominant star formation channel for a given galaxy.

We note that the intrinsic scatter predicted by our model systematically exceeds estimates of the scatter obtained using a variety of different 
empirical techniques that connect observed galaxies with the predicted halo population. Some examples of constraints on the scatter that have 
been reported include work using group catalogues \cite[$0.17 \, \mathrm{dex}$,][]{Yang09}, satellite kinematics \cite[$0.16 \, \mathrm{dex}$,][]{More09}, 
a combination of clustering, abundances and lensing \cite[$\approx 0.2 \, \mathrm{dex}$,][]{Leauthaud12}, clustering and group catalogues 
\cite[$0.2, 0.17 \, \mathrm{dex}$,][]{Reddick13,Rodriguez13}, and a combination of lensing and clustering \cite[$0.2 \, \mathrm{dex}$,][]{Zu15}. To take a specific
example, \cite{Reddick13} combine subhalo abundance matching with clustering and conditional stellar
mass function (estimated using a galaxy group catalogue) constraints to infer the scatter of the SHM relation in the local Universe. They rule out a
scatter as large as is predicted by our reference model and find that the intrinsic scatter is not a strong function of stellar mass, which is also in tension
with our reference model.

Regarding the tension associated with the amount of mass dependence in the scatter, this could indicate that, either we have overestimated the role of nuclear 
starbursts in the global stellar mass assembly process, 
or that the efficiency of SNe feedback should be constant at a given halo mass, irrespective of whether star formation is taking place
within discs or spheroids. We note that the latter possibility is assumed in the L-galaxies model \cite[][]{Guo11, Henriques13}, and 
that the SHM relation in that model does not predict a bimodal feature in the SHM relation at 
$M_{\mathrm{H}} = 10^{12.2} \, \mathrm{M_\odot}$ as a consequence \cite[see figure 5 in ][]{Contreras15}. This result can be reproduced
in \galform by simply changing the SNe feedback mass loading parametrisation to depend on halo circular velocity. We note that even with this
halo circular velocity dependent SNe feedback efficiency, the resulting SHM scatter ($\sigma \approx 0.3 \, \mathrm{dex}$) predicted by our model
is still in excess of the typical $\sigma \approx 0.2 \, \mathrm{dex}$ value estimated from applying empirical models to observational data.
We defer any further exploration of this issue to future work.

%% file: appendix_msmh2.tex
\section{Criteria for a non-evolving SHM relation for star forming galaxies}
\label{ap:msmh2}

Here, we explore the conditions required for a non-evolving SHM relation for star forming galaxies, based on the simplified 
analytical results presented in Section~\ref{sfg_msmh_invariance_section}. There we assumed that the instantaneous star formation
efficiency, $\eta_{\mathrm{SF}} \equiv \dot{M}_\star / (f_{\mathrm{B}} \dot{M}_\mathrm{H})$, scaled as $\beta_{\mathrm{ml}}^{-1}$, where 
$\beta_{\mathrm{ml}}$ is the mass loading factor for SNe feedback. By also assuming that the disc circular velocity scales with the halo 
circular velocity for haloes hosting star forming galaxies, we arrived at the following relation:

\begin{equation}
\eta_{\mathrm{SF}} \propto M_{\mathrm{H}}^{\alpha_{\mathrm{hot}} /3} \, \left[\bar{\rho}_{\mathrm{H}}(a)\right]^{\alpha_{\mathrm{hot}} /6}.
\label{ap_simple_scaling}
\end{equation}

\noindent For a non-evolving SHM relation, $\eta_{\mathrm{SF}}$ should be constant at a fixed halo mass. Eqn.~\ref{ap_simple_scaling}
contradicts this requirement because the mean halo density, $\bar{\rho}_{\mathrm{H}}(a)$, depends on expansion factor, independent of
halo mass. Integrating Eqn.~\ref{ap_simple_scaling} will therefore yield an evolving SHM relation.

In Section~\ref{sfg_msmh_invariance_section}, we circumvented this problem by assuming that $\bar{\rho}_{\mathrm{H}}(a)$ was constant
with expansion factor. In this case, integrating Eqn.~\ref{ap_simple_scaling} yields

\begin{equation}
M_{\star} \propto M_{\mathrm{H}}^{1 + \alpha_{\mathrm{hot}}/3} \left[\bar{\rho}_{\mathrm{H}}\right]^{\alpha_{\mathrm{hot}} /6}
\label{shm_nonevolv_req}
\end{equation}

\noindent where, if $\bar{\rho}_{\mathrm{H}}$ is regarded as constant, we arrive at the non-evolving SHM relation given by Eqn.~\ref{non_evolv_shm_eqn}.
To test the regimes where this assumption is valid, we can invert the process of integrating Eqn.~\ref{ap_simple_scaling} into Eqn.~\ref{shm_nonevolv_req},
differentiating Eqn.~\ref{shm_nonevolv_req} to give

\begin{eqnarray*}
\dot{M}_{\star} \propto & \left(1 + \frac{\alpha_{\mathrm{hot}}}{3}\right) M_{\mathrm{H}}^{\alpha_{\mathrm{hot}}/3} \dot{M}_{\mathrm{H}} \left[\bar{\rho}_{\mathrm{H}}\right]^{\alpha_{\mathrm{hot}} /6} \\
                       &+ \frac{\alpha_{\mathrm{hot}}}{6} M_{\mathrm{H}}^{1 + \alpha_{\mathrm{hot}}/3} \dot{\bar{\rho}}_{\mathrm{H}} \left[\bar{\rho}_{\mathrm{H}}\right]^{\alpha_{\mathrm{hot}} /6 -1}.
\label{invert_integral_shm}
\end{eqnarray*}

\noindent In order for this to be equivalent to Eqn.~\ref{ap_simple_scaling}, we require that

\begin{equation}
\frac{|\dot{M}_{\mathrm{H}}|}{M_{\mathrm{H}}} \gg F(a) \equiv \frac{\alpha_{\mathrm{hot}}}{2(3+\alpha_{\mathrm{hot}})} \frac{|\dot{\bar{\rho}}_{\mathrm{H}}(a)|}{\bar{\rho}_{\mathrm{H}}(a)}.
\label{shm_equality}
\end{equation}

\noindent In other words, given Eqn.~\ref{ap_simple_scaling}, a non-evolving SHM relation requires that haloes are growing
faster in mass than the rate with which they are changing in density. This inequality will not be satisfied in general. However,
it may be satisfied for haloes of a particular mass over some redshift range.

\begin{figure}
\includegraphics[width=20pc]{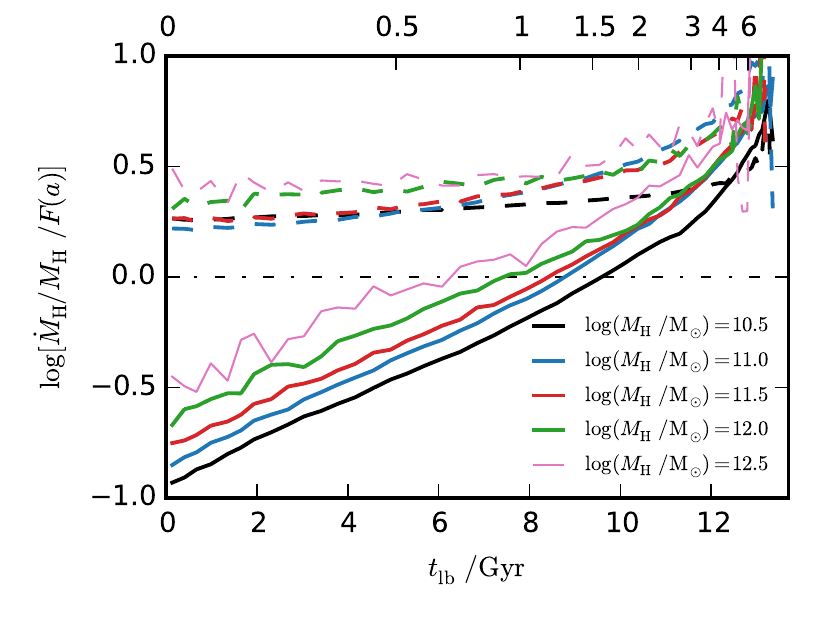}
\caption{Scaled ratio of average halo accretion rates to the rate with which the mean halo density, $\bar{\rho}_{\mathrm{H}}$, is changing with time, plotted as a function of lookback time.
Average halo formation rates are taken from our reference model for haloes that host star forming central galaxies at a given redshift.
The scaled rate of change in mean halo density, $F(a)$, is defined by \protect Eqn.~\ref{shm_equality}.
Solid lines show $\dot{M}_{\mathrm{H}}/M_{\mathrm{H}}/F(a)$ when $F(a)$ is calculated using mean halo densities evaluated from the spherical collapse model.
Dashed lines show$\dot{M}_{\mathrm{H}}/M_{\mathrm{H}}/F(a)$ when $F(a)$ is calculated by averaging over the mean halo densities taken directly from \galform.
Each coloured line corresponds to a different halo mass bin, as labelled.
The black dash-dotted horizontal line shows the line of equality, above which the condition for a non-evolving SHM relation is met for star forming galaxies.
}
\label{shm_equality_test}
\end{figure}

In Fig.~\ref{shm_equality_test}, we show, for a range of halo masses, the redshift range for which the inequality given by Eqn.~\ref{shm_equality} is satisfied. 
Here, we have selected haloes from our reference model that host star forming galaxies
at a given redshift, and then computed $\dot{M}_{\mathrm{H}}$, averaged over bins in halo mass. To calculate 
$|\dot{\bar{\rho}}_{\mathrm{H}}(a)|$, we consider both $\bar{\rho}_{\mathrm{H}}(a)$ calculated using the spherical collapse
model (solid lines) and calculated directly using halo circular velocities taken from our reference model (dashed lines).

Starting with halo densities computed from the spherical collapse model, Fig.~\ref{shm_equality_test} shows that that the 
SHM relation should be non-evolving for star forming galaxies when $z>1$. The exact redshift
where Eqn.~\ref{shm_equality} is met depends on halo mass, such that the inequality is met over a wider redshift range for more 
massive haloes.

Conversely, from Fig.~\ref{shm_equality_test}, we also expect that the SHM relation should evolve at lower redshifts ($z<1$)
if halo densities are computed using the spherical collapse model. However, significant
evolution is not seen in the SHM relation over this redshift range (in the halo mass range associated with star forming galaxies) for our reference model 
in, for example, Fig.~\ref{smhm_model_comp}. This can partially be explained by noting that for $z<1$, 
star formation rates and halo mass accretion rates at a given halo mass have dropped dramatically relative to higher redshifts. 

However, another very important consideration is that in our reference model, halo circular velocities (and hence the mean densities of haloes 
at fixed halo mass) of individual haloes are only updated when haloes double in mass.
While halo formation (mass doubling) events are very frequent at high redshift when halo mass accretion rates are very large, they become very
infrequent at low redshifts, for which halo mass accretion rates have dropped dramatically. Consequently, the average halo density
for haloes from our reference model will evolve more slowly with time than if the halo densities followed exactly the spherical 
collapse model. This effect can be seen directly by considering the dashed lines in Fig.~\ref{shm_equality_test}, which show
$F(M_{\mathrm{H}},a)$ evaluated from the average of halo densities taken directly from our reference model. In this case, it
is apparent that the inequality given by Eqn.~\ref{shm_equality}, on average, is met for all haloes over all redshifts. This helps to explain 
why the SHM model does not evolve significantly in our reference model.

%% file: appendix_msmh4.tex
\section{AGN feedback and the SHM break mass}
\label{ap:msmh4}

In Section~\ref{agn_fb_section}, we derived a simple expectation for how a threshold halo mass for AGN feedback to
be effective in suppressing cooling in hydrostatic haloes would evolve with cosmic time. To do so, we evaluated
a criterion for quasi-hydrostatic equilibrium by equating the cooling time at the mean gas density within a
halo to the freefall time at the halo virial radius. Here, we present a more detailed derivation
of this threshold halo mass, this time assuming a simple isothermal sphere density profile to evaluate Eqn.~\ref{agn_fb_eqn} at the cooling
radius. We also explore the role of the secondary criterion for effective AGN feedback in our model, which is that the AGN power must
be sufficient to offset radiative cooling from a quasi-hydrostatic halo.

\subsection{Isothermal sphere derivation}

As described in Section~\ref{agn_fb_section}, for AGN feedback to be effective at shutting down cooling we require
the cooling time, $t_{\mathrm{cool}}$, to exceed the the freefall time, $t_{\mathrm{ff}}$, by a factor $\alpha_{\mathrm{cool}}^{-1}$,

\begin{equation}
t_{\mathrm{cool}}(r=r_{\mathrm{cool}}) \geq t_{\mathrm{ff}}(r=r_{\mathrm{cool}}) / \alpha_{\mathrm{cool}}.
\label{agn_condition}
\end{equation}

\noindent We want to find a threshold halo mass, $M_{\mathrm{H}}(z)$, where this condition is met. To do this we first need to compute $r_{\mathrm{cool}}$ in terms of $t_{\mathrm{cool}}$.
The cooling time scales as

\begin{equation}
t_{\mathrm{cool}}(r=r_{\mathrm{cool}}) \propto \frac{T}{\rho(r=r_{\mathrm{cool}}) \Lambda_{\mathrm{cool}}(T,Z_{\mathrm{g}})},
\label{tcool_scaling}
\end{equation}

\noindent and the virial temperature scales as

\begin{equation}
T \propto V_{\mathrm{H}}^2 \propto \frac{M_{\mathrm{H}}}{r_{\mathrm{H}}}.
\end{equation}

\noindent If we assume that the mass distribution within a halo follows an isothermal sphere profile (truncated at the virial radius, $r_{\mathrm{H}}$)
such that 

\begin{equation}
\rho(r) \propto \frac{M_{\mathrm{H}}}{r_{\mathrm{H}} \, r^2},
\label{isothermal}
\end{equation}

\noindent then we can evaluate Eqn.~\ref{tcool_scaling}, yielding

\begin{equation}
t_{\mathrm{cool}}(r=r_{\mathrm{cool}}) \propto \frac{r_{\mathrm{cool}}^2}{\Lambda_{\mathrm{cool}}(T,Z_{\mathrm{g}})}.
\end{equation}

\noindent Rearranging, we find that the cooling radius scales as

\begin{equation}
r_{\mathrm{cool}}  \propto \sqrt{t_{\mathrm{cool}} \Lambda_{\mathrm{cool}}(T,Z_{\mathrm{g}})}.
\end{equation}

\noindent To evaluate this expression, we need a value for $t_{\mathrm{cool}}$. In \galform, this is the time since the
last halo formation event, $t_{\mathrm{form}}$. For the halo merger trees extracted from the {\sc{MR7}}\xspace
simulation, we find this can be well described by

\begin{equation}
t_{\mathrm{form}} \propto t_{\mathrm{H}} M_{\mathrm{H}}^{0.05},
\end{equation}

\noindent where $t_{\mathrm{H}} = 1/H(t)$ is the Hubble time.

For an isothermal sphere, the freefall time from a radius, $r$, (which must be within the virial radius) scales as

\begin{equation}
t_{\mathrm{ff}}(r) \propto \frac{r}{V_{\mathrm{H}}} \propto \frac{r}{M_{\mathrm{H}}^{1/3} \bar{\rho}_{\mathrm{H}}^{1/6}}. 
\end{equation}

\noindent We now have everything required to evaluate Eqn.~\ref{agn_condition},

\begin{equation}
t_{\mathrm{H}} M_{\mathrm{H}}^{0.05} \propto \frac{r_{\mathrm{cool}}}{M_\mathrm{H}^{1/3} \bar{\rho}_{\mathrm{H}}^{1/6}}.
\end{equation}

\noindent Substituting for $r_{\mathrm{cool}}$ yields

\begin{equation}
t_{\mathrm{H}} M_{\mathrm{H}}^{0.05} \propto \frac{\sqrt{t_{\mathrm{H}} M_{\mathrm{H}}^{0.05} \Lambda_{\mathrm{cool}}(T,Z_{\mathrm{g}})}}{M_\mathrm{H}^{1/3} \bar{\rho}_{\mathrm{H}}^{1/6}}.
\label{long}
\end{equation}

\noindent We also need to evaluate the mean halo density using the spherical collapse model

\begin{equation}
\bar{\rho}_{\mathrm{H}}(z) \propto \Delta_{\mathrm{c}}(z) \rho_{\mathrm{crit}}(z) \propto \Delta_{\mathrm{c}}(z) t_{\mathrm{H}}^{-2}.
\end{equation}

\noindent Eqn.~\ref{long} then reduces to

\begin{equation}
M_{\mathrm{H}} \propto \Lambda_{\mathrm{cool}}(T,Z_{\mathrm{g}})^{1.4} \, \Delta_{\mathrm{c}}(z)^{-0.47} \, t_{\mathrm{H}}^{-0.47}.
\label{mhalo_agn}
\end{equation}

\noindent This scaling is close to the simplified derivation presented in Section~\ref{agn_fb_section}, which, for
reference, yielded $M_{\mathrm{H}} \propto \Lambda_{\mathrm{cool}}(T,Z_{\mathrm{g}})^{1.5} \, \Delta_{\mathrm{c}}(z)^{1/4} \, t_{\mathrm{H}}^{-0.5}$.
The two derivations differ approximately by a factor of $\Delta_{\mathrm{c}}(z)^{3/4}$, which turns out to be unimportant relative to the
evolution in the cooling function, $\Lambda_{\mathrm{cool}}(T,Z_{\mathrm{g}})$.

\subsection{Hydrostatic equilibrium, AGN feedback and quenching}

\begin{figure}
\includegraphics[width=20pc]{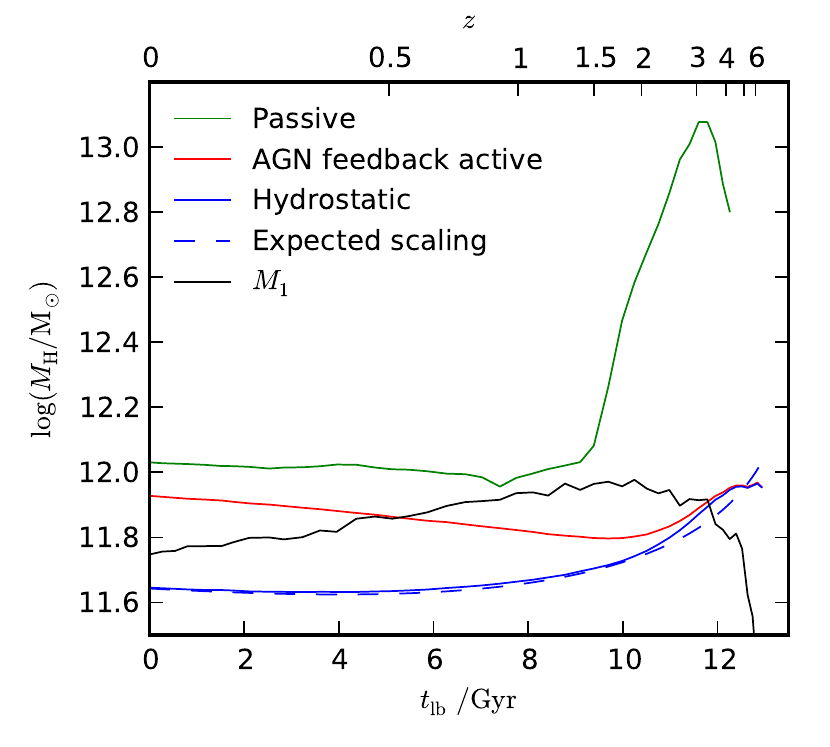}
\caption{Halo mass thresholds plotted as a function of lookback time for our reference model.
The green line shows the halo mass at which half of central galaxies are passive.
The red line shows the halo mass for which AGN feedback is actively suppressing cooling in half of the haloes hosting central galaxies.
The blue solid line shows the halo mass for which half of the haloes hosting central galaxies meet the quasi-hydrostatic 
equilibrium criterion for AGN feedback given by \protect Eqn.~\ref{agn_condition}.
The dashed blue line shows the redshift scaling given by \protect Eqn.~\ref{mhalo_agn}, normalised to agree with the solid blue line at $z=0$.
The solid black line shows the break halo mass in the median SHM relation, $M_{\mathrm{1}}$, for our reference model.}
\label{passive_mhalo_test}
\end{figure}

\noindent To check whether Eqn.~\ref{mhalo_agn} is a reasonable description of what occurs in \galform, we compute the halo mass where the
fraction of central galaxies that meet the quasi-hydrostatic equilibrium criterion given by Eqn.~\ref{agn_condition} is equal to 
$0.5$. The resulting halo mass (solid blue line) is compared to the expectation from Eqn.~\ref{mhalo_agn} (dashed blue line) in 
Fig.~\ref{passive_mhalo_test}. To evaluate Eqn.~\ref{mhalo_agn}, we take the median gas temperature and metallicity at the halo
mass given by the solid blue line to compute the evolution of the cooling function. From Fig.~\ref{passive_mhalo_test}, it is 
apparent that the halo mass where the hydrostatic criterion is met is essentially constant up to $z=2$ and then increases mildly for $z>2$.

For the AGN feedback model in \galform, a second requirement for AGN feedback to be effective in suppressing cooling is that
the maximum AGN power (taken to be a fraction of the Eddington luminosity of the black hole) is sufficient to balance the radiative
luminosity of the cooling flow \cite[][]{Bower06}. In practice, provided that the central galaxy hosts a central super-massive
black hole, this criterion is almost always met, such that the hydrostatic criterion given by Eqn.~\ref{agn_condition} controls
whether AGN feedback is effective in a given halo. However, there is a non-negligible fraction of central galaxies in our
model in haloes close the hydrostatic threshold mass given by Eqn.~\ref{mhalo_agn} that do not host super-massive black holes.
These are the model galaxies that have not undergone a gas rich merger or a disc instability over their lifetime, and typically
have bulge-to-total ratios $\approx 0$. As a consequence of this galaxy population, the threshold halo mass where AGN feedback is
active in suppressing cooling in half of the haloes hosting central galaxies (solid red line in Fig.~\ref{passive_mhalo_test}) 
is actually larger than the mass where the hydrostatic equilibrium criterion is met (solid blue line).

Another consideration is that once AGN feedback becomes active in a given halo at suppressing gas inflow onto the central galaxy,
there can still be an appreciable delay before star formation shuts down in that galaxy. The length of the delay depends primarily
on the strength of SNe feedback in ejecting cold gas from the central galaxy. The solid green line in Fig.~\ref{passive_mhalo_test}
shows the halo mass above which half of the central galaxies are passive. Below $z=2$, this mass closely traces (with an offset) the
halo mass where AGN feedback is active. However, for $z>2$, the halo mass above which central galaxies are typically quenched increases
strongly with redshift. This strong evolution for $z>2$ is not reflected by a corresponding evolution in the SHM break mass, 
$M_{\mathrm{1}}$, which we overplot in Fig.~\ref{passive_mhalo_test} (solid black line). We attribute this difference primarily
to the fact that the width of the sigmoid function that describes the passive fraction of central galaxies as a function of halo mass
is significantly larger at high redshift compared to low redshift. As such, at high redshift, there is a non-negligible population 
of passive central galaxies at significantly lower halo masses than is indicated by the red line in Fig.~\ref{passive_mhalo_test}
(which shows the halo mass where the sigmoid function is equal to $0.5$).

\label{lastpage}